\documentclass[a4paper,11pt]{article}
\usepackage{jheppub} 

\usepackage[english]{babel}

\usepackage{bbold}

\usepackage[utf8]{inputenc}
\usepackage{graphicx}
\usepackage{amssymb}
\usepackage{amsmath}
\usepackage{xcolor}
\usepackage{hyperref}
\usepackage{multirow}
\usepackage{slashed}
\usepackage{subcaption}
\usepackage{bm}

\newcommand{\<}{\langle}
\renewcommand{\>}{\rangle}
\newcommand{\be}{\begin{equation}}
\newcommand{\ee}{\end{equation}}
\newcommand{\ba}{\begin{eqnarray}}
\newcommand{\ea}{\end{eqnarray}}
\newcommand{\la}{\label}
\newcommand{\calD}{\ensuremath{\mathcal{D}}}
\newcommand{\calbD}{\ensuremath{\bar{\mathcal{D}}}}

\newcommand{\deriv}[2]{\ensuremath{\frac{\mathrm{d} #1}{\mathrm{d} #2}}}

\preprint{CERN-TH-2025-239, MITP-25-074}

\title{Euclidean coordinate-space perturbation theory\\ with a single mass scale}

\author[a]{Christoph L.\ Schr\"oder,}

\affiliation[a]{PRISMA$^+$ Cluster of Excellence \& Institut f\"ur Kernphysik,
Johannes Gutenberg-Universit\"at Mainz,
D-55099 Mainz, Germany}
\emailAdd{cschroe@students.uni-mainz.de}

\author[a,b,c]{Harvey~B.\ Meyer}

\affiliation[b]{Helmholtz~Institut~Mainz,
Staudingerweg 18, D-55128 Mainz, Germany}

\affiliation[c]{
Theoretical Physics Department, CERN, 1211 Geneva 23, Switzerland
}

\emailAdd{h.b.meyer@cern.ch}

 \date{November 2025}

\begin{document}

\abstract{We develop elements of coordinate-space perturbation theory for massive quantum field theories in general $d$-dimensional Euclidean space. Using the expansion in Gegenbauer polynomials, we provide analytic expressions for several three-point correlation functions in theories with one massive and one massless field.
To this end, a class of antiderivatives of products of two Bessel functions multiplied by a power of their common argument are studied systematically.
We expect these results to be useful in
perturbative calculations involving vertices of high degree, at finite temperature and/or in finite volume, as well as in auxiliary perturbative computations for a lattice QCD treatment of hadronic effects in precision observables.
As an illustration, we compute the one-loop coordinate-space propagator of a massive particle coupled to a massless one, both in the vacuum and at finite temperature.}

\maketitle

\section{Introduction}

In a seminal paper published in 1980, Chetyrkin, Kataev and Tkachov developed the Gegenbauer-polynomial coordinate-space technique~\cite{Chetyrkin:1980pr} for computing dimensionally regularized massless Feynman diagrams.
In this paper, we proceed to generalize the technique to diagrams with massive propagators. More precisely, most of the examples considered in the following consist of a number of propagators with mass $m$ and others that are massless, a type of kinematics we refer to as `QED-like'. A necessary ingredient in these calculations, the Gegenbauer expansion of the massive scalar propagator was already given in~\cite{Groote:2005ay}.
As compared to the original publication~\cite{Chetyrkin:1980pr}, in which multi-loop momentum-space amplitudes are computed with coordinate-space techniques, we focus more strongly on intrinsically coordinate-space amplitudes. 
As a typical application, the Gegenbauer-polynomial technique has been used~\cite{Asmussen:2022oql} in $d=4$ to compute the QED part of the diagrams yielding the hadronic light-by-light contribution to the muon $(g-2)$.
A second difference in emphasis is that with the massive propagators being Bessel functions, the radial integrals become non-trivial, and we will devote a substantial portion of this article to computing precisely such integrals.

Textbooks on quantum field theory often begin with formulating Feynman rules in position-space perturbation theory, before moving swiftly to the corresponding momentum-space rules. There is a simple reason for that. Recall that in position space, the integration variables correspond to the coordinates of vertices\footnote{Note that, due to translation invariance, the naive number of required position-space integrals is one less than the number of vertices.}, whereas in momentum space they are the loop momenta. Now, for a connected graph made of $n$-legged vertices, we have the following relation (see e.g.~\cite{Weinberg:1995mt}) between the number of loops $L$, the number of vertices $V$ and the number of external legs $E$:
\begin{equation}
    (n-2) V = 2(L-1) + E.
\end{equation}
For $n=3$, for instance, as in QED and for most vertices in the Standard Model, at high orders the number of vertices grows twice as fast as the number of loops. For contact interactions, $n=4$, the growth rate breaks even, and higher-order vertices favour the position-space approach in terms of the naive number of integrals to be performed. Thus position-space methods have potential in effective field theories with contact interactions involving four or more legs per vertex. Increasing the number of external legs carrying definite momenta, on the other hand, clearly favours the momentum-space methods, as this corresponds to computing Fourier integrals for vertices attached to external legs.
Of course, these simple arguments do not account for the possibility that integrals may factorize or otherwise simplify. In fact, we shall see instances of position-space amplitudes that can be computed and re-used as building blocks in later calculations. In momentum space, too, certain amplitudes are re-usable in more extended diagrams, the simplest example being the self-energy correction to a propagator.

Besides applications in field theories with contact interactions and for evaluating the electroweak parts of Feynman diagrams involving non-perturbative `QCD blobs' to be computed in lattice QCD, perturbative position-space methods have been known for a long time to be useful in calculations performed in a finite volume~\cite{Collins:1984xc,Luscher:1985dn}, the thermal boundary conditions of the Matsubara formalism being an important special case~\cite{Gasser:1987zq}. In high-order perturbative calculations of the QCD free energy, the frequency/spatial-coordinates representation was found to be particularly well suited~\cite{Arnold:1994ps,Arnold:1994eb}. Indeed the `vacuum diagrams' to be computed, having no external legs ($E=0$), lend themselves naturally to a coordinate-space treatment.

Before proceeding, we briefly mention a few  directions tangential to the present work.
(a) Calculations in coordinate space have been used to give a different perspective on the renormalization of a quantum field theory~\cite{Freedman:1991tk}. In particular, a central idea in this approach is to give a precise meaning to products of distributions.
(b) Coordinate-space techniques have also found applications in other regularizations. They have been used in lattice perturbation theory calculations~\cite{Luscher:1985wf} as a way to reduce the number of numerical sums/integrals to be computed.
(c) The Gegenbauer-polynomial technique has been considered in $d$-dimensional momentum space~\cite{Terrano:1980af}.
For $d=4$, the propagator $((p-q)^2+m^2)^{-1}$ has a relatively simple expansion in the relevant polynomials $C_n^{(1)}(\hat p\cdot \hat q)$,
a fact exploited already in~\cite{Rosner:1967zz}, when dimensional regularization had not been discovered yet.

\begin{figure}[t]
    \centering
    \includegraphics[width=0.32\linewidth]{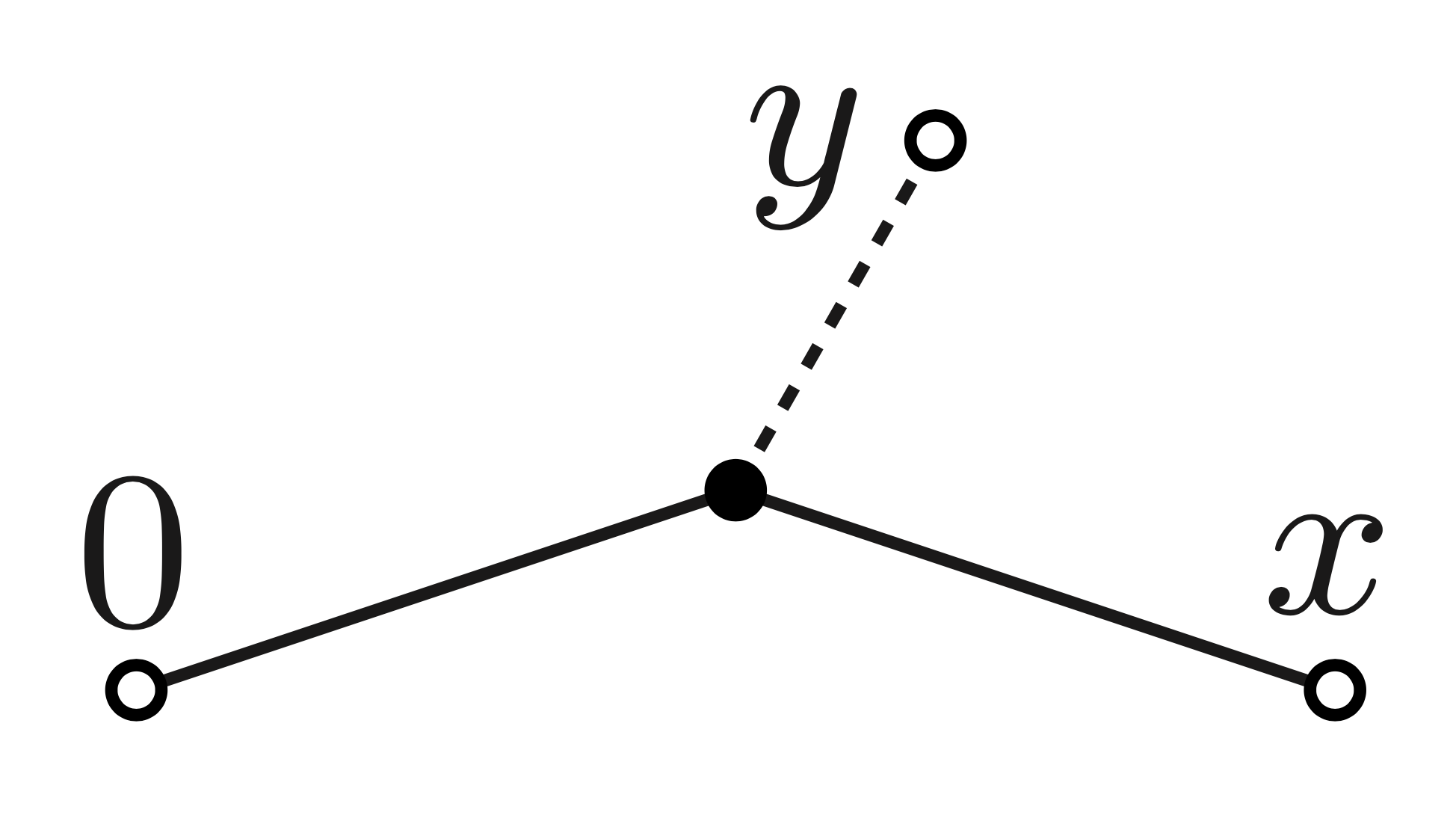}
    \caption{The coordinate-space three-point function. Full (dashed) lines represent massive (massless) propagators.}
    \label{fig:3ptFct}
\end{figure}

Consider then for concreteness a Euclidean quantum field theory with a massive complex scalar field $\phi(x)$ and a massless real scalar field $\chi(x)$, interacting via the Yukawa-like Lagrangian term ${\cal L}_{\rm int}=g\chi \phi^*\phi$. Let  $G_m(x)$ denote the propagator for a scalar particle of mass $m$.
An important building block in coordinate-space perturbation theory is the three-point function illustrated at leading order (LO) in $g$ in Fig.\ \ref{fig:3ptFct}. Its analytic expression reads
\begin{equation}\label{eq:3ptdef}
\< \phi(x) \chi(y) \phi^*(0) \> \stackrel{\rm LO}{=} 
-g\int d^dz\; G_m(z)\, G_{0}(y-z) \,G_m(x-z).
\end{equation}
Applying the Gegenbauer polynomial expansion method to perform the angular integrals leads to radial integrals of the form ($n\in\mathbb{N}_0$, $\lambda=(d-2)/2$)
\begin{align}
    & \int dz  \,z^{n+1} \,K_\lambda(mz)\,  {\cal D}_{\lambda+n}(mz),
    \\
    & \int dz \, \frac{1}{z^{2\lambda+n-1}}  \,K_\lambda(mz)\,  {\cal D}_{\lambda+n}(mz),
\end{align}
where ${\cal D}\in\{I,K\}$ stands for a modified Bessel function, either of the first or the second kind. The integrals are needed for arbitrary upper or lower bound, in other words the full anti-derivative is needed.
This leads us to study systematically certain classes of anti-derivatives, such as 
\begin{equation}
    \int dz  \,z^{\ell} \,{\cal D}_\lambda(mz)\,  \bar{\cal D}_{\lambda+n}(mz),
\end{equation}
with $n,\ell\in \mathbb{N}_0$ and ${\cal D},\bar{\cal D}\in\{I,K\}$. We find a simple condition under which the result is expressible as a sum of terms of the same type as the integrand itself, i.e.\ containing a power of their argument and a product of two  modified Bessel functions or none at all.

The goal of this paper is to illustrate the computational methods and to provide expressions for a class of indefinite integrals, as well as for several re-usable amplitudes that can be considered to be building blocks for coordinate-space perturbation theory.
We begin in section~\ref{sec:elements} with a presentation of the basic elements of coordinate-space theory in general $d$-dimensional Euclidean space.
The bulk of the mathematical results, which concern antiderivatives of products of two modified Bessel functions, are collected in section~\ref{sec:antideriv}.
We then consider three field-theoretic calculations: the leading-order coordinate-space three-point function (section~\ref{sec:3ptFct}), the two-point function at one-loop order (section~\ref{sec:2pt1loop}), and the mixed momentum-coordinate space three-point function (section~\ref{sec:3ptmised}). An application to scalar field theory at finite temperature is given in section~\ref{sec:finiteT} before we collect our conclusions in section~\ref{sec:concl}.

\section{Elements of coordinate-space perturbation theory\label{sec:elements}}

We work in $d=2\lambda+2$ dimensions.
The volume of the $S_{d-1}$ unit sphere, which has codimension one in $d$ dimensions, is given by
\begin{equation}
\Omega_d = \frac{2\pi^{d/2}}{\Gamma(d/2)} = \frac{2\pi^{\lambda+1}}{\Gamma(\lambda+1)}\;.
\end{equation}
The average of a function $f$ of $x\in\mathbb{R}^d$ over the unit sphere is defined as 
\begin{equation}
  \<  f(x) \>_{\hat x} \equiv \frac{1}{\Omega_d} \int d\Omega_{\hat x}\; f(x),
\end{equation}
where $d\Omega_{\hat x}$ is the solid angle element associated with point $\hat x$ on the unit sphere, where $\hat x = x/r$ with $r=|x|$. For instance, for $d=3$ we have $d\Omega_{\hat x}=\sin\theta \,d\theta \,d\phi$, where $(r,\theta,\phi)$ are the spherical coordinates of point $x$.

\subsection{Propagators}
The massive propagator solves the  
`screened Poisson equation'
\begin{equation}
    (-\triangle +m^2)\, G_m^{(\lambda)}(x) = \delta^{(d)}(x),
    \end{equation}
   with $\triangle=\partial_\mu \partial_\mu$ denoting the $d$-dimensional Laplacian and $\delta^{(d)}(x)$ the Dirac delta distribution.
  Explicitly, the propagator reads
\begin{equation}
G_m^{(\lambda)}(x) = \frac{m^\lambda}{(2\pi)^{\lambda+1}}\; \frac{K_\lambda(m|x|)}{|x|^\lambda}\,.
   \end{equation}
   In the limit $m\to0$, one obtains the massless propagator
\begin{equation}
G_0^{(\lambda)}(x) 
= \frac{\Gamma(\lambda)}{4 \,\pi^{\lambda+1}|x|^{2\lambda}} \,.
\end{equation}
   
Propagators in $d$ and $(d+2)$ dimensions are related through differentiation,
\begin{equation}\label{eq:Gxderiv}
- \frac{d}{d(x^2)} G_m^{(\lambda)}(x) = \pi\, G_m^{(\lambda+1)}(x).
\end{equation}
Differentiating with respect to the mass, instead, lowers the index $\lambda$ by one unit,
\begin{equation}\label{eq:Gmderiv}
-  \frac{d}{d(m^2)} G_m^{(\lambda)}(x) = \frac{1}{4\pi} G_m^{(\lambda-1)}(x).
\end{equation}
Exploiting the fact that $K_{-\lambda}(z) = K_{\lambda}(z)$, we also note the formal property
\begin{equation}
G_m^{(-\lambda)}(x) =  \frac{1}{2\pi} \left(\frac{2\pi |x|}{m}\right)^\lambda K_\lambda(m|x|) = (2\pi)^{2\lambda}  G_{|x|}^{(\lambda)}(m\,\hat\epsilon)\,,
\end{equation}
with $\hat\epsilon\in \mathbb{R}^d$ an arbitrary unit vector, which interchanges the role of the mass and the distance of propagation.

The propagator of a Dirac fermion reads
\begin{eqnarray}\label{eq:Sm1}
S^{(\lambda)}_m(x) &=&  (-\partial_\mu^{(x)}\gamma_\mu + m) G_m^{(\lambda)}(x) 
\\ &=&  x_\mu\gamma_\mu \,2\pi\, G_m^{(\lambda+1)}(x) + m G_m^{(\lambda)}(x).
\label{eq:Sm2}
\end{eqnarray}

\subsection{Convolution of two propagators}

\begin{figure}
     \centering
     \includegraphics[width=0.5\linewidth]{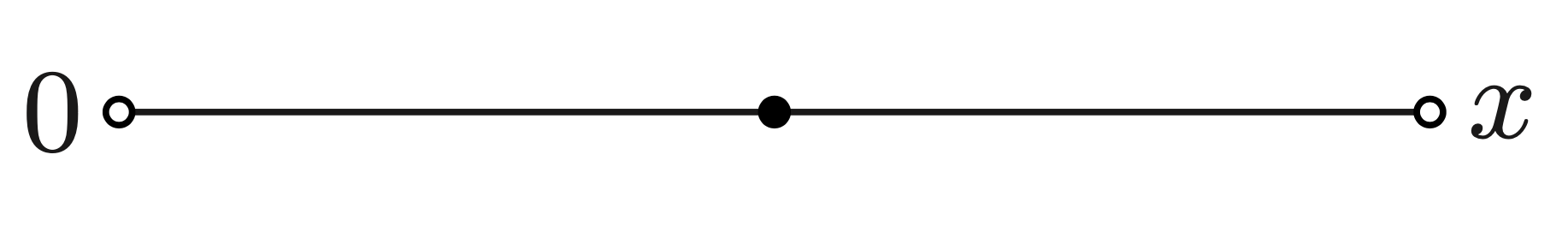}
     \caption{Feynman diagram corresponding to the convolution of two coordinate-space propagators.}
     \label{fig:convol}
 \end{figure}

Here we recall an elementary result on 
the convolution of two propagators, illustrated by the diagram in Fig.\ \ref{fig:convol}. It is most easily obtained via the detour into Fourier space, since the convolution then becomes a simple product and we can decompose the latter according to 
\begin{equation}
    \frac{1}{k^2+m^2}\cdot \frac{1}{k^2+M^2}
    = \frac{1}{M^2-m^2} \left(\frac{1}{k^2+m^2}-\frac{1}{k^2+M^2}  \right).
\end{equation}
Thus one finds
\begin{equation}\label{eq:GmStarGM}
 (G^{(\lambda)}_m *  G_M^{(\lambda)})(x) \equiv  \int d^dy\; G^{(\lambda)}_m(x-y) G^{(\lambda)}_M(y) = 
    \frac{1}{M^2-m^2} \left(G^{(\lambda)}_m(x) - G^{(\lambda)}_M(x) \right).
\end{equation}
The convolution can be obtained for $M=m$ as a limiting case. Using Eq.\ (\ref{eq:Gmderiv}) in the last step, one finds
\begin{equation}\label{eq:GmStarGm}
  (G^{(\lambda)}_m *  G^{(\lambda)}_m)(x) 
  = -\frac{\partial}{\partial m^2}G^{(\lambda)}_m(x)
  = \frac{1}{4\pi} G^{(\lambda-1)}_m(x).  
\end{equation}
An alternative way to derive this result is to note that in a free massive scalar field theory, 
\[-\frac{\partial}{\partial m^2} G_m^{(\lambda)}(x) = 
-\frac{\partial}{\partial m^2} \<\phi(x)\phi^*(0)\> = \int d^dy\,\left\<\phi(x) \phi^*(y)\phi(y)\,\phi^*(0)\right\>_c = (G^{(\lambda)}_m *  G^{(\lambda)}_m)(x) . \]

\subsection{Gegenbauer polynomials}

The Gegenbauer polynomials can be defined via the expansion of a generating function,
\begin{equation}
\frac{1}{(1-2xt+t^2)^\lambda} = \sum_{n=0}^\infty C_n^{\lambda}(x)\,t^n.
\end{equation}
The first three polynomials read
$C_0^\lambda(x) = 1$,
$C_1^\lambda(x) = 2\lambda\,x$ and 
$C_2^\lambda(x) = \lambda(2(\lambda+1)x^2 -1 )$.
The lower index of the polynomial determines whether it is even or odd,
\begin{eqnarray}
C_n^\lambda(-x) = (-1)^n C_n^\lambda(x),
\end{eqnarray}
and its overall normalization is such that 
\begin{equation}
    C_n^\lambda(1) = \frac{(2\lambda)_n }{ n!}\;,
\end{equation}
where $(a)_n \equiv \Gamma(a+n)/\Gamma(a)$ is the Pochhammer symbol.
In the present context, the most useful property of this set of polynomials is their orthogonality on the $S_{d-1}$ unit sphere in $d=2\lambda+2$ dimensions,
\begin{eqnarray}
\Big\< C_n^{\lambda}(\hat x\cdot\hat u)\;C_\ell^{\lambda}(\hat y\cdot \hat u)\Big\>_{\hat u} 
&=& \frac{\lambda}{n+\lambda}\,\delta_{n\ell}\; C_n^{\lambda}(\hat x\cdot \hat y).
\label{eq:CnOrthog}
\end{eqnarray}
For $\lambda = 1$ these polynomials are known as Chebyshev polynomials of the second kind $U_n$.

The Gegenbauer expansion of the massive propagator  reads~\cite{Groote:2005ay}
\begin{eqnarray}
G_m^{(\lambda)}(x-y) &=& \sum_{n=0}^\infty \gamma^{(m,\lambda)}_n(|x|,|y|)\; C^{(\lambda)}_n(\hat x\cdot \hat y)
\\
\gamma^{(m,\lambda)}_n(|x|,|y|) &=& \frac{\Gamma(\lambda)(\lambda+n) }{2\,\pi^{\lambda+1} (|x||y|)^\lambda}
 \Big(\theta(|x|-|y|)  \,I_{\lambda+n}(m|y|)\, K_{\lambda+n}(m|x|) \label{eq:elements:gamma_n^m}
\\ && \qquad \qquad \qquad 
+ \theta(|y|-|x|)  \,I_{\lambda+n}(m|x|) \,K_{\lambda+n}(m|y|) \Big),
\nonumber 
\end{eqnarray}
a result that goes back to a contribution of the year 1900 by Macdonald according to \mbox{Watson's} classic treatise on Bessel functions~\cite{Watson}.
An important aspect of the expansion coefficients is that for $|x|>|y|$, the expression factorizes in a function of $|x|$ times a function of $|y|$, and similarly for the opposite ordering.
As a special case, the expansion coefficients for the massless propagator read~\cite{Chetyrkin:1980pr}
\begin{eqnarray}
\gamma^{(0,\lambda)}_n(|x|,|y|) &=& \frac{\Gamma(\lambda)}{4 \,\pi^{\lambda+1}}
\left[\theta(|x|-|y|) \frac{1}{|x|^{2\lambda}} \left(\frac{|y|}{|x|}\right)^n
+ \theta(|y|-|x|) \frac{1}{|y|^{2\lambda}} \left(\frac{|x|}{|y|}\right)^n\right].
\end{eqnarray}
We also note that, from the relation (\ref{eq:GmStarGm}) between propagators in $d$ and $(d+2)$ dimensions, 
 the following expansion is immediately obtained,
\ba
G_m^{(\lambda-1)}(x-y) &=& \sum_{n=0}^\infty \Big(-4\pi \frac{\partial}{\partial m^2} \gamma_n^{(m,\lambda)}(|x|,|y|)\Big) \,C_n^{(\lambda)}(\hat x\cdot \hat y).
\ea
An alternative derivation of the coefficients of $G_m^{(\lambda-1)}(x-y)$ in the polynomials $C_n^{(\lambda)}(\hat x\cdot \hat y)$ proceeds via the relation between Gegenbauer polynomials corresponding to $d$ and $d-2$ dimensions (polynomials of negative degree are defined to vanish here),
\begin{align}
	C_n^{(\lambda-1)}(x) &= 
   \frac{\lambda-1}{\lambda+n-1} \left( C_n^{(\lambda)}(x) - C_{n-2}^{(\lambda)}(x)\right).
\end{align}
In this way, one finds the relation
\begin{align}\label{eq:elements:dm2gamma_n^mB}
    -4\pi \frac{\partial}{\partial m^2} \gamma_n^{(m,\lambda)}(|x|,|y|)= 
    (\lambda-1)\left(
    \frac{\gamma_n^{(m,\lambda-1)}(|x|,|y|)}{\lambda+n-1} - \frac{\gamma_{n+2}^{(m,\lambda-1)}(|x|,|y|)}{\lambda+n+1}\right).
\end{align}
 
    Finally, the following Gegenbauer-polynomial expansion is useful for Fourier transformations,
\begin{equation}\label{eq:eipxGg}
e^{ik\cdot x} = 2^\lambda\Gamma(\lambda) \sum_{n=0}^\infty (\lambda+n) i^n \frac{J_{\lambda+n}(|k||x|)}{(|k||x|)^\lambda}
C^{(\lambda)}_n(\hat k\cdot \hat x).
\end{equation}

\section{Antiderivatives of products of two modified Bessel functions\label{sec:antideriv}}
In the preceding section, we reviewed the fact that in coordinate-space, the massive scalar propagator is mainly given by a modified Bessel function (of the second kind), and the propagator with a difference of two vectors as argument can be written as an expansion with modified Bessel functions in its coefficients. Hence, when evaluating Feynman diagrams in coordinate space, the usual integrations over the vertices, post applying the Gegenbauer expansion and the orthogonality relation of the corresponding polynomials, yield radial integrals with integrands consisting of products of a monomial and two modified Bessel functions. Because the expression for the expansion coefficients depends on which input vector is larger (cf.\ Eq.\ \eqref{eq:elements:gamma_n^m}), the radial integrals have variable integration bounds, i.e.\ they amount to indefinite integrals. It is hence worthwhile investigating this class of antiderivatives thoroughly, which we will do in this section. In the case of integrals appearing in calculations performed in $d=4$ dimensions, a systematic study was presented in~\cite{BachelorThesisSchroder}.

\subsection{Known results due to Lommel, et al.}

In this subsection, we summarize a class of known results for the antiderivatives in question. From Watson \cite{Watson}, two important results attributed to Lommel exist. We have transcribed them from ordinary to modified Bessel function. Watson refers to the first as \textit{Lommel's first method}, which we shall simply call \textit{Lommel 1}. For modified Bessel functions, it states that
\begin{align}\label{eq:lom1KI}
&\int \frac{I_\mu(z) K_\nu(z)}{z} \,dz = z \frac{I_{\mu+1}(z) K_\nu(z) - I_\mu(z) K_{\nu+1}(z)}{\mu^2 - \nu^2} + \frac{I_\mu(z) K_\nu(z)}{\mu + \nu},\\
&\int \frac{I_\mu(z) K_\nu(z)}{z} \,dz = \frac{z}{\mu^2 - \nu^2} \left(I_\mu'(z) K_\nu(z) - I_\mu(z) K_\nu'(z)\right),
\label{eq:lom1KIb}
\end{align}
\begin{align}\label{eq:lom1II}
&\int \frac{I_\mu(z) I_\nu(z)}{z} \,dz = z \frac{I_{\mu+1}(z) I_\nu(z) - I_\mu(z) I_{\nu+1}(z)}{\mu^2 - \nu^2} + \frac{I_\mu(z) I_\nu(z)}{\mu + \nu},\\
&\int \frac{I_\mu(z) I_\nu(z)}{z} \,dz = \frac{z}{\mu^2 - \nu^2} \left(I_\mu'(z) I_\nu(z) - I_\mu(z) I_\nu'(z)\right),
\label{eq:lom1IIb}
\\  \label{eq:lom1KK}
&\int \frac{K_\mu(z) K_\nu(z)}{z} \,dz = -z \frac{K_{\mu+1}(z) K_\nu(z) - K_\mu(z) K_{\nu+1}(z)}{\mu^2 - \nu^2} + \frac{K_\mu(z) K_\nu(z)}{\mu + \nu},\\
&\int \frac{K_\mu(z) K_\nu(z)}{z} \,dz = \frac{z}{\mu^2 - \nu^2} \left(K_\mu'(z) K_\nu(z) - K_\mu(z) K_\nu'(z)\right).
\label{eq:lom1KKb}
\end{align}
For equal orders one could evaluate the limit e.g.\ by L'H\^{o}pital's rule, but since this involves derivatives with respect to the order, we will see a  simpler way to come to a result in a moment.

Watson also includes what he calls \textit{Lommel's second method}, which we will refer to as \textit{Lommel 2}, and this comprises integral recursion relations that read
\begin{align}\label{eq:lom2recII}
(\rho + \mu + \nu) \int z^{\rho-1} I_\mu(z) I_\nu(z) \,dz &- (\rho - \mu - \nu -2) \int z^{\rho-1} I_{\mu+1}(z) I_{\nu+1}(z) \,dz \nonumber\\
&= z^\rho \left(I_\mu(z) I_\nu(z) - I_{\mu+1}(z) I_{\nu+1}(z)\right),
\end{align}
\begin{align}\label{eq:lom2recKK}
(\rho + \mu + \nu) \int z^{\rho-1} K_\mu(z) K_\nu(z) \,dz &- (\rho - \mu - \nu - 2) \int z^{\rho-1} K_{\mu+1}(z) K_{\nu+1}(z) \,dz \nonumber\\
&= z^\rho \left(K_\mu(z) K_\nu(z) - K_{\mu+1}(z) K_{\nu+1}(z)\right),
\end{align}
and finally, for the mixed case,
\begin{align}\label{eq:lom2recKI}
(\rho + \mu + \nu) \int z^{\rho-1} K_\mu(z) I_\nu(z) \,dz &+ (\rho - \mu - \nu - 2) \int z^{\rho-1} K_{\mu+1}(z) I_{\nu+1}(z) \,dz \nonumber\\
&= z^\rho \left(K_\mu(z) I_\nu(z) + K_{\mu+1}(z) I_{\nu+1}(z)\right).
\end{align}
One observes that all these recursion relations collapse in the following two cases, and then present the solution to one antiderivative: $\rho = \mu + \nu + 2$ and $\rho = -\mu - \nu$. For example, for the mixed case this gives the results
\begin{equation}\label{eq:lom2:speca}
2(\mu + \nu + 1) \int z^{\mu+\nu+1} K_\mu(z) I_\nu(z) \,dz = z^{\mu+\nu+2} \left(K_\mu(z) I_\nu(z) + K_{\mu+1}(z) I_{\nu+1}(z)\right),
\end{equation}
and,
\begin{equation}\label{eq:lom2:specb}
2(\mu + \nu - 1) \int K_\mu(z) I_\nu(z) \frac{dz}{z^{\mu + \nu - 1}} = - \frac{1}{z^{\mu + \nu - 2}} \left(K_{\mu-1}(z) I_{\nu-1}(z) + K_\mu(z) I_\nu(z)\right).
\end{equation}
Similarly, results can be found for the other two combinations of Bessel functions.

Also from Watson, one may now deduce integrals of the type $\int z^{-1} \mathcal{D}_p(z) \bar{\mathcal{D}}_p(z) \,dz$ in an elegant way, where $p$ is a non-zero integer. These are
\begin{equation}\label{eq:lom:ell-1_equal_int_order_II}
\int I_p^2(z) \frac{dz}{z} = \frac{1}{2p} \left\{I_p^2(z) + (-1)^p \left[-I_0^2(z) - 2 \sum_{k=1}^p (-1)^k I_k^2(z)\right]\right\},
\end{equation}
\begin{equation}
\int K_p^2(z) \frac{dz}{z} = \frac{1}{2p} \left\{K_p^2(z) + (-1)^p \left[-K_0^2(z) - 2 \sum_{k=1}^p (-1)^k K_k^2(z)\right]\right\},
\end{equation}
and finally,
\begin{equation}
\int K_p(z) I_p(z) \frac{dz}{z} = \frac{1}{2p} \left\{-K_0(z) I_0(z) - 2 \sum_{k=1}^p K_k(z) I_k(z) + K_p(z) I_p(z)\right\}.
\end{equation}
It shall be noted that for non-integer $p$ we at least have a result (cf.  \cite{Watson}) for functions of the first kind, namely
\begin{equation}\label{eq:lom:ell-1_equal_order}
2\nu \int_0^z \frac{I_\nu^2(t)}{t} dt = I_\nu^2(z) + 2 \sum_{n=1}^\infty (-1)^n I_{\nu+n}^2(z),
\end{equation}
but there seems to be no simple result for the indefinite integrals $\int z^{-1} \calD_0(z) \calbD_0(z) \,dz$.

\subsection{Extension of Schafheitlin's recursion relation and further special results}\label{sec:schafheitlin}

In the following, we provide an extended version of a recursion relation, which in its simpler form is originally due to Schafheitlin, among indefinite integrals of products of Bessel functions. 
In contrast to Lommel's recursion relation it involves a recursion with respect to the monomial instead of the order of Bessel functions.

First, consider Bessel's modified differential equation,
\begin{equation}
z^2 y''(z) + z y'(z) - (z^2 + \nu^2)y = 0,
\end{equation}
fulfilled by both modified cylinder functions $\mathcal{D}_\nu$, $\bar{\mathcal{D}}_\nu$. From this,
\begin{align}
\int z^\rho &(z^2 + \nu^2) \mathcal{D}_\nu(z) \bar{\mathcal{D}}_\mu(z) \, dz \nonumber\\
&= \int z^\rho \left\{z^2 \mathcal{D}_\nu''(z) + z \mathcal{D}_\nu'(z)\right\} \bar{\mathcal{D}}_\mu(z) \, dz,
\end{align}
where the differential equation was applied on $\mathcal{D}_\nu$. Then, from a partial integration, we get
\begin{align}
\int z^\rho &(z^2 + \nu^2) \mathcal{D}_\nu(z) \bar{\mathcal{D}}_\mu(z) \,dz \nonumber\\
&= z^{\rho+2} \mathcal{D}_\nu'(z) \bar{\mathcal{D}}_\mu(z) - (\rho + 1) \int z^{\rho+1} \mathcal{D}_\nu'(z) \bar{\mathcal{D}}_\mu(z) \,dz - \int z^{\rho+2} \mathcal{D}_\nu'(z) \bar{\mathcal{D}}_\mu'(z) \,dz.
\end{align}
Now let us repeat the process, letting the differential equation act on $\bar{\mathcal{D}}_\mu$ this time:
\begin{align}
\int z^\rho &(z^2 + \mu^2) \mathcal{D}_\nu(z) \bar{\mathcal{D}}_\mu(z) \,dz \nonumber\\
&= z^{\rho+2} \mathcal{D}_\nu(z) \bar{\mathcal{D}}_\mu'(z) - (\rho + 1) \int z^{\rho+1} \mathcal{D}_\nu(z) \bar{\mathcal{D}}_\mu'(z) \,dz - \int z^{\rho+2} \mathcal{D}_\nu'(z) \bar{\mathcal{D}}_\mu'(z) \,dz.
\end{align}
Symmetrizing both expressions yields
\begin{align}
\mathcal{I} &\equiv (\rho + 1) \left\{\int z^\rho (z^2 + \nu^2) \mathcal{D}_\nu(z) \bar{\mathcal{D}}_\mu(z) \,dz + \int z^\rho (z^2 + \mu^2) \mathcal{D}_\nu(z) \bar{\mathcal{D}}_\mu(z) \,dz\right\} \label{eq:IschafDef}\\
&= (\rho + 1) z^{\rho+2} S_{\mu\nu}(z) - (\rho + 1)^2 \int z^{\rho+1} S_{\mu\nu}(z)\,dz - 2 (\rho + 1) \int z^{\rho+2} \mathcal{D}_\nu'(z) \bar{\mathcal{D}}_\mu'(z) \,dz, \label{eq:schaf:I}
\end{align}
where the symmetric combination is defined as
\begin{equation}
S_{\mu\nu}(z) := \mathcal{D}_\nu'(z) \bar{\mathcal{D}}_\mu(z) + \mathcal{D}_\nu(z) \bar{\mathcal{D}}_\mu'(z)
=  \deriv{}{z} \Bigl(\mathcal{D}_\nu(z) \bar{\mathcal{D}}_\mu(z)\Bigr).
\end{equation}

We will now consider the last integral appearing in Eq.\ (\ref{eq:schaf:I}) and perform integration by parts. We obtain
\begin{align}
(\rho + 3) \int z^{\rho+2} &\mathcal{D}_\nu'(z) \bar{\mathcal{D}}_\mu'(z) \,dz \nonumber\\
&= z^{\rho+3} \mathcal{D}_\nu'(z) \bar{\mathcal{D}}_\mu'(z) - \left\{\int z^{\rho+3} \mathcal{D}_\nu''(z) \bar{\mathcal{D}}_\mu'(z) \,dz + \int z^{\rho+3} \mathcal{D}_\nu'(z) \bar{\mathcal{D}}_\mu''(z) \,dz\right\}.
\end{align}
From the differential equation obeyed by the modified Bessel functions, we infer
\begin{align}
z^2 \mathcal{D}_\nu''(z) = (z^2 + \nu^2) \mathcal{D}_\nu(z) - z \mathcal{D}_\nu'(z), \\
z^2 \bar{\mathcal{D}}_\mu''(z) = (z^2 + \mu^2) \bar{\mathcal{D}}_\mu(z) - z \bar{\mathcal{D}}_\mu'(z),
\end{align}
so that by rearranging it follows that 
\begin{align}
(\rho &+ 1) \int z^{\rho+2} \mathcal{D}_\nu'(z) \bar{\mathcal{D}}_\mu'(z) \,dz \nonumber\\
&= z^{\rho+3} \mathcal{D}_\nu'(z) \bar{\mathcal{D}}_\mu'(z) - \int z^{\rho+3} S_{\mu\nu}(z) \,dz - \int z^{\rho+1} \left[\mu^2 \mathcal{D}_\nu'(z) \bar{\mathcal{D}}_\mu(z) + \nu^2 \mathcal{D}_\nu(z) \bar{\mathcal{D}}_\mu'(z) \right] \,dz.
\end{align}

The integrals involving the symmetric combination are easily handled.
Performing integration by parts gives
\begin{equation} \label{eq:schaf:pi_S_rho+1}
\int z^{\rho+1} S_{\mu\nu}(z) \,dz = z^{\rho+1} \mathcal{D}_\nu(z) \bar{\mathcal{D}}_\mu(z) - (\rho + 1) \int z^\rho \mathcal{D}_\nu(z) \bar{\mathcal{D}}_\mu(z)\, dz
\end{equation}
and similarly for $\rho$ replaced by $(\rho+2)$.
Hence we obtain for the expression in Eq.\ (\ref{eq:schaf:I}) the form 
\begin{align}
\mathcal{I} = (\rho + 1) &z^{\rho + 2} S_{\mu\nu}(z) - (\rho + 1)^2 z^{\rho+1} \mathcal{D}_\nu(z) \bar{\mathcal{D}}_\mu(z) + 2 z^{\rho+3} \mathcal{D}_\nu(z) \bar{\mathcal{D}}_\mu(z) - 2 z^{\rho+3} \mathcal{D}_\nu'(z) \bar{\mathcal{D}}_\mu'(z) \nonumber\\
&+ (\rho + 1)^3 \int z^\rho \mathcal{D}_\nu(z) \bar{\mathcal{D}}_\mu(z) \,dz
- 2(\rho + 3) \int z^{\rho+2} \mathcal{D}_\nu(z) \bar{\mathcal{D}}_\mu(z) \,dz \nonumber\\
&+ 2 \int z^{\rho+1} \left[\mu^2 \mathcal{D}_\nu'(z) \bar{\mathcal{D}}_\mu(z) + \nu^2 \mathcal{D}_\nu(z) \bar{\mathcal{D}}_\mu'(z) \right] \,dz. \label{eq:schaf:I2}
\end{align}
On the other hand we have from the definition of ${\cal I}$ in Eq.\ (\ref{eq:IschafDef}) that
\begin{equation} \label{eq:schaf:I1}
\mathcal{I} = 2 (\rho + 1) \int z^{\rho+2} \mathcal{D}_\nu(z) \bar{\mathcal{D}}_\mu(z) \,dz + (\rho + 1) (\mu^2 + \nu^2) \int z^\rho \mathcal{D}_\nu(z) \bar{\mathcal{D}}_\mu(z) \,dz,
\end{equation}
so that by equating both \eqref{eq:schaf:I2} and \eqref{eq:schaf:I1} as in \eqref{eq:schaf:I} we obtain our `milestone' relation,
\begin{align}
4 (\rho + 2) &\int z^{\rho+2} \mathcal{D}_\nu(z) \bar{\mathcal{D}}_\mu(z) \,dz + (\rho + 1) \left[\mu^2 + \nu^2 - (\rho + 1)^2\right] \int z^\rho \mathcal{D}_\nu(z) \bar{\mathcal{D}}_\mu(z) \,dz \nonumber\\
&= \{\star\} + 2 \int z^{\rho + 1} \left[\mu^2 \mathcal{D}_\nu'(z) \bar{\mathcal{D}}_\mu(z) + \nu^2 \mathcal{D}_\nu(z) \bar{\mathcal{D}}_\mu'(z) \right] \,dz \label{eq:schaf:gen}
\end{align}
where
\begin{equation} \label{eq:schaf:gen_star}
\{\star\} = (\rho + 1) z^{\rho+2} S_{\mu\nu}(z) - (\rho + 1)^2 z^{\rho+1} \mathcal{D}_\nu(z) \bar{\mathcal{D}}_\mu(z) + 2 z^{\rho+3} \mathcal{D}_\nu(z) \bar{\mathcal{D}}_\mu(z) - 2 z^{\rho+3} \mathcal{D}_\nu'(z) \bar{\mathcal{D}}_\mu'(z)
\end{equation}
no longer contains any integrals.

In the following, we simplify Eq.\ (\ref{eq:schaf:gen}) in several cases motivated by the need to handle the remaining integral on the right-hand side of the equation.
We can rewrite that integral in the following way,
\begin{align} \label{eq:schaf:2ints}
\int z^{\rho+1} &\left[\mu^2 \mathcal{D}_\nu'(z) \bar{\mathcal{D}}_\mu(z) + \nu^2 \mathcal{D}_\nu(z) \bar{\mathcal{D}}_\mu'(z) \right] \,dz \\
&= \frac{\mu^2+\nu^2}{2} \int z^{\rho+1} S_{\mu\nu}(z) \,dz + \frac{\mu^2-\nu^2}{2} \int z^{\rho+1} \left(\mathcal{D}_\nu'(z) \bar{\mathcal{D}}_\mu(z) - \mathcal{D}_\nu(z) \bar{\mathcal{D}}_\mu'(z)\right)\,dz. 
\nonumber
\end{align}

\paragraph{The case $\mu^2=\nu^2$}
In the special case $\mu^2 - \nu^2 = 0$, the last term of Eq.\ (\ref{eq:schaf:2ints}) vanishes. Using again \eqref{eq:schaf:pi_S_rho+1}, we obtain
\begin{align}
\int z^{\rho+1} &\left[\mu^2 \mathcal{D}_\nu'(z) \bar{\mathcal{D}}_\mu(z) + \nu^2 \mathcal{D}_\nu(z) \bar{\mathcal{D}}_\mu'(z) \right] \,dz = \nonumber\\
&\nu^2 \left\{z^{\rho+1} \mathcal{D}_\nu(z) \bar{\mathcal{D}}_\mu(z) - (\rho + 1) \int z^\rho \mathcal{D}_\nu(z) \bar{\mathcal{D}}_\mu(z)\, dz\right\}, \label{eq:schaf:2ints_first_sol}
\end{align}
and hence, with $\mu = \pm \nu$ the milestone relation Eq.\ (\ref{eq:schaf:gen}) becomes
\begin{equation}
4 (\rho + 2) \int z^{\rho+2} \mathcal{D}_\nu(z) \bar{\mathcal{D}}_\mu(z) \,dz + (\rho + 1) \left[4\nu^2 - (\rho + 1)^2\right] \int z^\rho \mathcal{D}_\nu(z) \bar{\mathcal{D}}_\mu(z) \,dz = \{\star\}_{\text{equal}}, \label{eq:schaf:equal}
\end{equation}
where
\begin{align} 
\{\star\}_{\text{equal}} &= 
 (\rho + 1) z^{\rho+2} S_{\mu\nu}(z) - \left[(\rho + 1)^2 - 2 \nu^2)\right] z^{\rho+1} \mathcal{D}_\nu(z) \bar{\mathcal{D}}_\mu(z) \nonumber\\
&\qquad + 2 z^{\rho+3} \mathcal{D}_\nu(z) \bar{\mathcal{D}}_\mu(z) 
- 2 z^{\rho+3} \mathcal{D}_\nu'(z) \bar{\mathcal{D}}_\mu'(z). \label{eq:schaf:equal_star}
\end{align}
Eq.\ (\ref{eq:schaf:equal}) amounts to a recursion relation in the degree of the monomial, for the particular case $\mu^2=\nu^2$.
If one were to replace the two modified Bessel functions by the same regular Bessel function, one would obtain the well-known recursion relation due to Schafheitlin, as found in Watson~\cite{Watson}, section 5.14.

\paragraph{The milestone relation in the case $\rho=-2$}
We briefly mention that in this case, one may immediately evaluate the second integral in Eq.\ \eqref{eq:schaf:2ints} by means of Lommel~1 or the appropriate formulae for equal integer order. Notice that the case where an integral $\int z^{-1} \calD_0(z) \calbD_0(z) \,dz$ occurs simply cancels. We return  to the $\rho=-2$ case in section \ref{sec:shaf_reloaded}.

\paragraph{The general case, excluding $\rho= -1$}
We can also make use of Lommel 1 here. In order to do so, notice that
\begin{align}
	\deriv{}{z} &\Bigl(\calbD_\mu(z) \calD_\nu'(z) - \calbD_\mu'(z) \calD_\nu(z)\Bigr) = (\nu^2 - \mu^2) \deriv{}{z} \Bigl(\frac{1}{z} \int \frac{\calbD_\mu(z) \calD_\nu(z)}{z} d z\Bigr) \nonumber\\
	&= (\nu^2-\mu^2) \Biggl\{-\frac{1}{z^2} \int \frac{\calbD_\mu(z) \calD_\nu(z)}{z} d z + \frac{1}{z} \frac{\calbD_\mu(z) \calD_\nu(z)}{z} \Biggr\} \nonumber\\
	&= -\frac{\calbD_\mu(z) \calD_\nu'(z) - \calbD_\mu'(z) \calD_\nu(z)}{z} + (\nu^2-\mu^2) \frac{\calbD_\mu(z) \calD_\nu(z)}{z^2}.
\end{align}
Then, upon integrating by parts,
\begin{align}
	(\rho+2) \int z^{\rho+1} &\Bigl(\calbD_\mu(z) \calD_\nu'(z) - \calbD_\mu'(z) \calD_\nu(z)\Bigr) \,d z	\nonumber\\
	&= z^{\rho+2} \Bigl(\calbD_\mu(z) \calD_\nu'(z) - \calbD_\mu'(z) \calD_\nu(z)\Bigr) \nonumber\\
	&\qquad- (\nu^2-\mu^2) \int z^\rho \calbD_\mu(z) \calD_\nu(z) \,d z \nonumber\\
	&\qquad+ \int z^{\rho+1} \Bigl(\calbD_\mu(z) \calD_\nu'(z) - \calbD_\mu'(z) \calD_\nu(z)\Bigr) \,d z,
\end{align}
we obtain after rearrangements
\begin{align}
	(\rho+1) &\int z^{\rho+1} \Bigl(\calbD_\mu(z) \calD_\nu'(z) - \calbD_\mu'(z) \calD_\nu(z)\Bigr) \,d z = \nonumber\\
	& z^{\rho+2} \Bigl(\calbD_\mu(z) \calD_\nu'(z) - \calbD_\mu'(z) \calD_\nu(z)\Bigr) - (\nu^2-\mu^2) \int z^\rho \calbD_\mu(z) \calD_\nu(z) \,d z.
\end{align}
This expression can be inserted into Eq.\ (\ref{eq:schaf:2ints}), in which the integral over $z^{\rho+1}S_{\mu\nu}(z)$ is handled once again using
Eq.\ (\ref{eq:schaf:pi_S_rho+1}).
We then insert the entire expression (\ref{eq:schaf:2ints}) into the milestone relation Eq.\ \eqref{eq:schaf:gen} and perform rearrangements, yielding for $\rho \ne -1$
\begin{align}
	4(\rho+2) &\int z^{\rho+2} \calD_\nu(z) \calbD_\mu(z) \,d z \nonumber\\
	&- \Biggl[\frac{(\mu^2-\nu^2)^2	}{\rho+1} + (\rho+1) \Bigl((\rho+1)^2 - 2(\mu^2+\nu^2)\Bigr)\Biggr] \int z^\rho \calD_\nu(z) \calbD_\mu(z) \,d z = \{\star\}_{\rho \ne -1}, \label{eq:schafheitlin_full}
\end{align}
where
\begin{align}
	\{\star\}_{\rho \ne -1} &= (\rho+1) z^{\rho+2} S_{\mu\nu}(z) - \bigl[(\rho+1)^2 - (\mu^2 + \nu^2)\bigr] z^{\rho+1} \calbD_\mu(z) \calD_\nu(z) \nonumber\\
	&\qquad- 2 z^{\rho+3} \Bigl(\calbD_\mu'(z) \calD_\nu'(z) - \calbD_\mu(z) \calD_\nu(z)\Bigr) \nonumber\\
	&\qquad+ \frac{\mu^2-\nu^2}{\rho+1} z^{\rho+2} \Bigl(\calbD_\mu(z) \calD_\nu'(z) - \calbD_\mu'(z) \calD_\nu(z)\Bigr).
    \label{eq:schafheitlin_fullB}
\end{align}
Eq.\ (\ref{eq:schafheitlin_full}) is the sought-after general recursion relation in the degree of the monomial, valid for $\rho\neq -1$, and is one of the main results of this section.

\paragraph{The special case $\rho=\sigma-1$, $\nu=\mu + \sigma$}
In this case, we obtain explicit results for a two-parameter class of integrals:
it is easily verified that the expression in square brackets in Eq.\ \eqref{eq:schafheitlin_full} vanishes, wherefrom we obtain
\begin{equation}\label{eq:schaf:spcl_II}
	2(\sigma+1) \int z^{\sigma+1} I_\mu(z) I_{\mu+\sigma}(z) \, d z = z^{\sigma+2} \Bigl(I_\mu(z) I_{\mu+\sigma}(z)-I_{\mu-1}(z) I_{\mu+\sigma+1}(z)\Bigr).
\end{equation}	
This result is a generalization, respectively modification of Lommel~1 and~2. 
Although our derivation needed to assume $\sigma\neq 0$, the antiderivatives given in this paragraph are valid for arbitrary $\sigma\in\mathbb{R}$, as one verifies by differentiation.
Notice that for the case of two modified Bessel functions of the second kind, one can easily retrieve the corresponding result directly from Lommel~2 by using $K_\mu = K_{-\mu}$. This gives
\begin{equation}\label{eq:schaf:spcl_KK}
	2(\sigma+1) \int z^{\sigma+1} K_\mu(z) K_{\mu+\sigma}(z) \, d z = z^{\sigma+2} \Bigl(K_\mu(z) K_{\mu+\sigma}(z)-K_{\mu-1}(z) K_{\mu+\sigma+1}(z)\Bigr).
\end{equation}	
For mixed products we obtain the following results,
\begin{align}
	2(\sigma+1) \int z^{\sigma+1} K_\mu(z) I_{\mu+\sigma}(z) \, d z &= z^{\sigma+2} \Bigl(K_\mu(z) I_{\mu+\sigma}(z)+ K_{\mu-1}(z) I_{\mu+\sigma+1}(z)\Bigr),\label{eq:schaf:spcl_KI}\\
	2(\sigma+1) \int z^{\sigma+1} I_\mu(z) K_{\mu+\sigma}(z) \, d z &= z^{\sigma+2} \Bigl(I_\mu(z) K_{\mu+\sigma}(z)+I_{\mu-1}(z) K_{\mu+\sigma+1}(z)\Bigr). \label{eq:schaf:spcl_IK}
\end{align}

\subsection{An algorithm for finding antiderivatives for non-negative integer powers}\label{sec:algpos}

Here we want to investigate integrals of the form 
\begin{equation}
	\int z^\ell \calD_\mu(z) \calbD_{\mu+n}(z) \,d z,	
\end{equation}
where $\ell$ and $n$ are non-negative integers, $\mu$ is a real number. Notice that by shifting $\mu$ by an integer, one can also incorporate expressions of the form $\int z^\ell \calD_\nu(z) \calbD_{\nu-n}(z) \,d z = \int z^\ell \calbD_\mu(z) \calD_{\mu+n}(z) \,d z$ on defining $\mu = \nu - n$.

\begin{figure}[ht]
	\centering
	\includegraphics[width=0.5\textwidth]{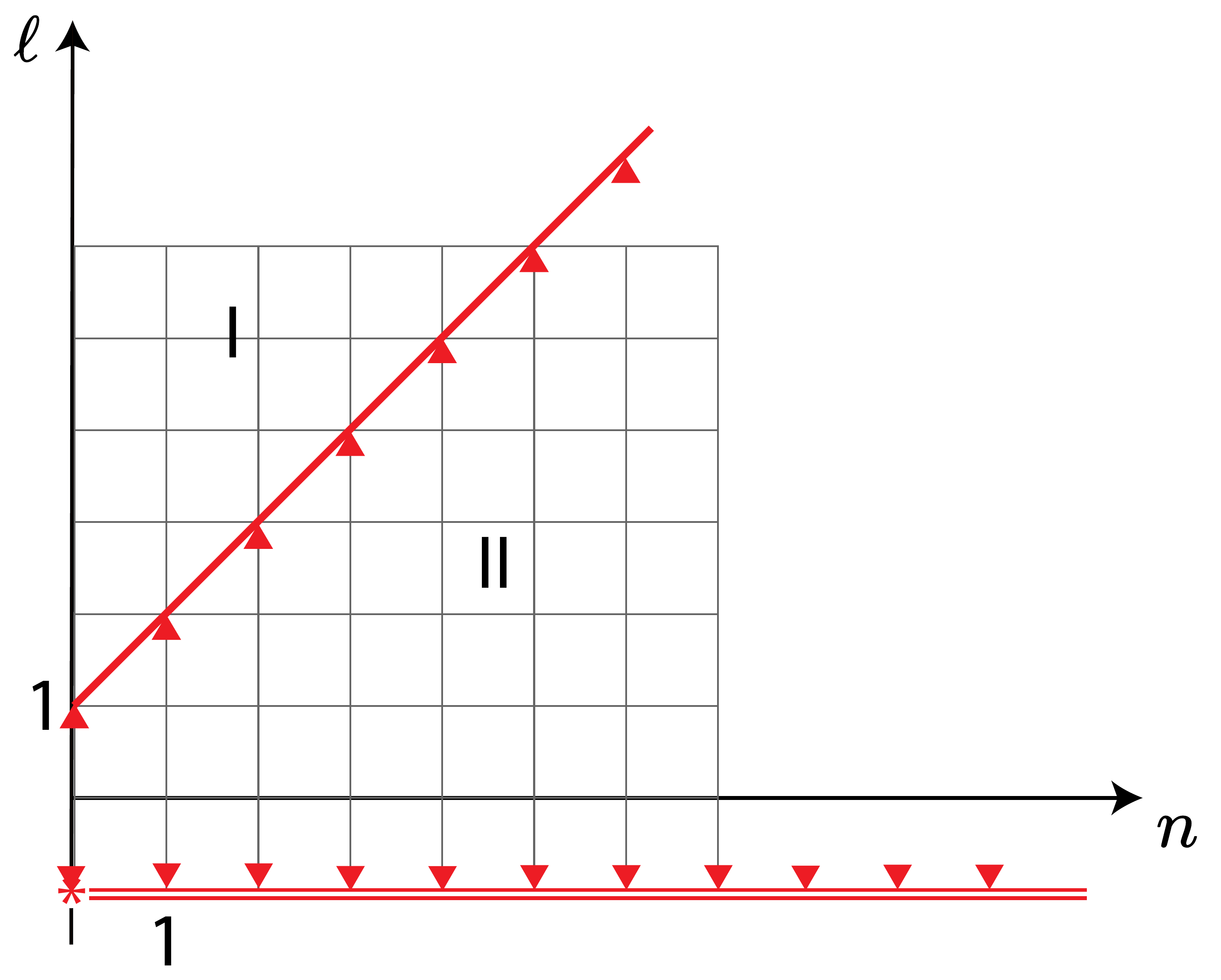}	
	\caption{Representation of the $\ell$-$n$-plane including the two shorelines where the result is known, see in particular Eq.\ \eqref{eq:schaf:spcl_II} and following, and Lommel~1. For the point ($n = 0,\ell = -1)$, indicated by the star, consider Eqs.\ \eqref{eq:lom:ell-1_equal_int_order_II} and following. Two regions are defined, I above the $\ell = n+1$-magic-line, and II below this line. 
    Applying the recursion relation Eq.\ (\ref{eq:schafheitlin_full}) corresponds to taking a two-unit vertical step.
    The triangles signal that the $\ell=-1$ lines cannot be reached by taking a downward vertical step, and the $\ell=n+1$ line cannot be reached by taking an upward vertical step.}\label{fig:algpos:regions}
\end{figure}

Let us investigate the integral expression in the $\ell$-$n$-plane. We have two straight lines for which we know an easy expression for the antiderivative (from now on referred to as shorelines). One is given by the formulae of the last section, namely eqs. \eqref{eq:schaf:spcl_II} to \eqref{eq:schaf:spcl_IK}, and there the result is known for $\ell = n+1$. The other known result applies to the case $\ell = -1$ (cf.\ Lommel 1). When both orders of Bessel functions are equal (i.e.\ $n = 0$), one can simplify this according to eq. \eqref{eq:lom:ell-1_equal_order}, which in the case of non-zero integer $\mu$ further simplifies to a finite summation, see the equations above eq. \eqref{eq:lom:ell-1_equal_order}. Only for $\mu = 0$ and $n = 0$ there seems to be no easy expression. This leads to a diagram as in Fig.\ \ref{fig:algpos:regions}, in which we have defined two regions, I and II.

We will now show that the existence of a simple form for the antiderivative in question depends on the parity of $\ell + n$. By simple we mean an expression that is of the same form as the integrand: a (finite) sum of a monomial multiplied by two modified Bessel functions. To see this, we develop an algorithmic way to obtain the antiderivative. Pictorially spoken, we will show how to reach only the shorelines from any point in the grid of the $\ell$-$n$-plane if $\ell + n$ is odd for any non-zero real $\mu$. In the case that $\mu$ is a non-zero integer or we start from region I, only finite summations are involved.

To develop our algorithm, let us investigate the allowed ways to move in the grid in the $\ell$-$n$-plane. Firstly, there is the recursion relation of Schafheitlin developed in the last section. Here, the power of the monomial is shifted by two while the orders of Bessel functions remain unchanged. There are two important caveats with this relation: First, it is not valid to go from power $+1$ to $-1$, hence one cannot reach the $\ell=-1$-magic-line using this relation. Secondly, the relation collapses when approaching the $\ell=n+1$-magic-line from below, hence, such a step is also forbidden.

Secondly, there is the usual recursion relation for the Bessel functions. This can be used to create a sum of two terms, one with the same power but the $n$ changed by two, and the other with both the power and the $n$ changed by one unit. Thus, this almost always creates two distinct strands that have to be followed. The exception is the case $n = 1$, where one can derive a recursion with only one strand,
\begin{equation}\label{eq:algpos:rec0_II}
	\int z^s I_\delta(z) I_{\delta+1}(z) \,d z = -\left(\delta + \frac{s}{2}\right) \int z^{s-1} I_\delta^2(z) \,d z + \frac{z^s}{2} I_\delta^2(z),
\end{equation}
valid for real $s$ and $\delta$.

We start the proof by writing
\begin{equation}
    \int z^s I_\delta(z) I_{\delta+1}(z) \,d z = \int z^{s-1} I_\delta(z) \Bigl(z I_{\delta+1}(z)\Bigr)\,d z.	
\end{equation}
Then, we can use the well-known identity $z I_\delta'(z) = z I_{\delta+1}(z) + \delta I_\delta(z)$ to obtain
\begin{equation}
    \int z^s I_\delta(z) I_{\delta+1}(z) \,d z = \int z^{s} I_\delta(z) I_\delta'(z)\,d z - \delta \int z^{s-1} I_\delta^2(z) \,d z.	
\end{equation}
For the first integral, use integration by parts, which yields
\begin{equation}
    \int z^{s} I_\delta(z) I_\delta'(z)\,d z = z^s I_\delta^2(z) - \int \Bigl(s z^{s-1} I_\delta^2(z) + z^s I_\delta'(z) I_\delta(z)\Bigr) \,d z,
\end{equation}
or on rearrangement
\begin{equation}
    \int z^{s} I_\delta(z) I_\delta'(z)\,d z = \frac{z^s}{2} I_\delta^2(z) - \frac{s}{2} \int z^{s-1} I_\delta^2(z) \,d z.
\end{equation}
Hence the final result is obtained on insertion.

\begin{figure}[t]
\centering
\begin{subfigure}[c]{0.45\textwidth}
	\centering
	\includegraphics[width=\textwidth]{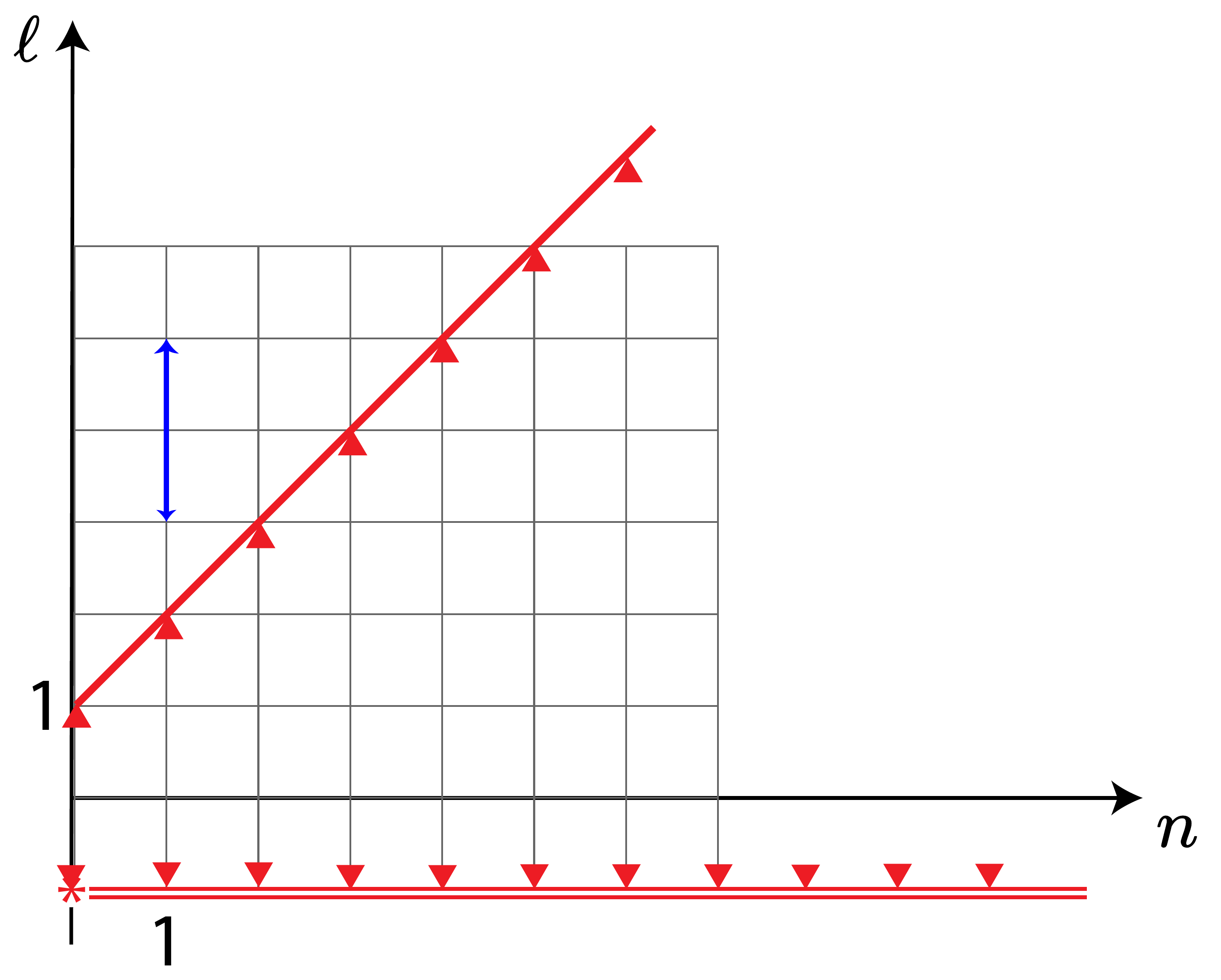}	
\end{subfigure}
\begin{subfigure}[c]{0.45\textwidth}
	\centering
	\includegraphics[width=\textwidth]{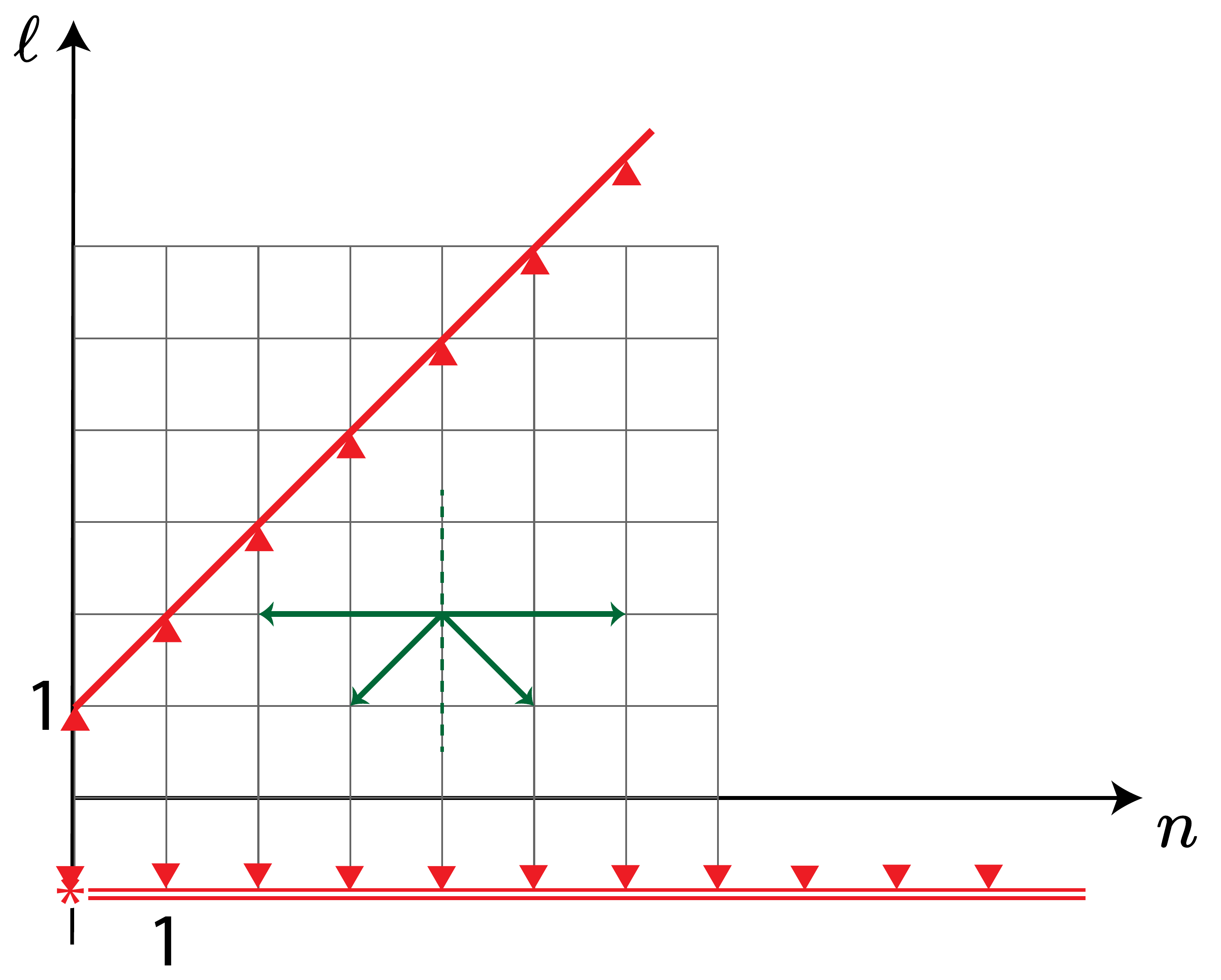}	
\end{subfigure}
\begin{subfigure}[c]{0.45\textwidth}
	\centering
	\includegraphics[width=\textwidth]{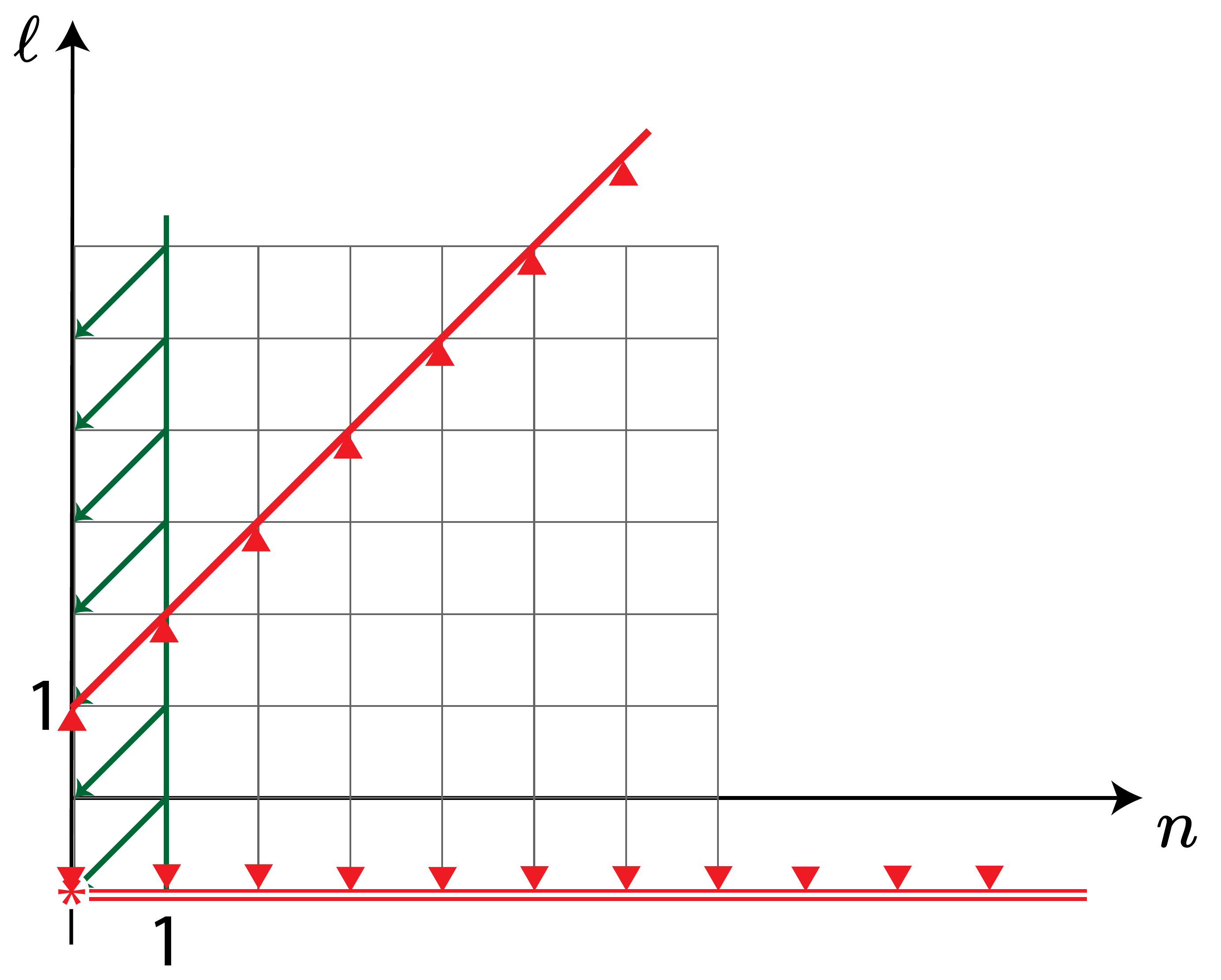}	
\end{subfigure}
\caption{Allowed operations to be taken in the $\ell$-$n$-plane represented by arrows. As before, the shorelines are drawn in red. Starting from the top right and going clockwise: our extended integral recursion relation originally due to Schafheitlin, see Eq.\ (\ref{eq:schafheitlin_full}) (see the main text for caveats), the usual recursion relation for the modified Bessel function, which always creates two strands which either both move to the left or both to the right, and finally the special recursion relation only valid when $n=1$; see Eq.\ \eqref{eq:algpos:rec0_II}.}
\label{fig:algpos:steps}
\end{figure}

There are similar formulae which can be derived in the same way for two modified Bessel functions of the second kind and for mixed products, in particular
\begin{equation}\label{eq:algpos:rec0_KK}
	\int z^s K_\delta(z) K_{\delta+1}(z) \,d z = \left(\delta + \frac{s}{2}\right) \int z^{s-1} K_\delta^2(z) \,d z - \frac{z^s}{2} K_\delta^2(z),
\end{equation}
and when using the fact that the Wronskian of $K_\delta(z)$ and $I_\delta(z)$ is given by $1/z$, one obtains also
\begin{align}
	\int z^s K_\delta(z) I_{\delta+1}(z) \,d z &= -\left(\delta + \frac{s}{2}\right) \int z^{s-1} K_\delta(z) I_\delta(z) \,d z + \frac{z^s}{2} K_\delta(z) I_\delta(z) \nonumber\\
    &\qquad+ \frac{1}{2} \begin{cases}\log |z|, &s = 0,\\z^s/s, &s \ne 0,\end{cases} \label{eq:algpos:rec0_KI}
\end{align}
and
\begin{align}
	\int z^s I_\delta(z) K_{\delta+1}(z) \,d z &= \left(\delta + \frac{s}{2}\right) \int z^{s-1} I_\delta(z) K_\delta(z) \,d z - \frac{z^s}{2} I_\delta(z) K_\delta(z) \nonumber\\
    &\qquad+ \frac{1}{2} \begin{cases}\log |z|, &s = 0,\\z^s/s, &s \ne 0.\end{cases} \label{eq:algpos:rec0_IK}
\end{align}

The three possible steps, described in the last paragraphs, to move in the grid in the $\ell$-$n$-plane are graphically summarized in fig. \ref{fig:algpos:steps}.

\begin{figure}[ht]
	\centering
	\includegraphics[width=0.5\textwidth]{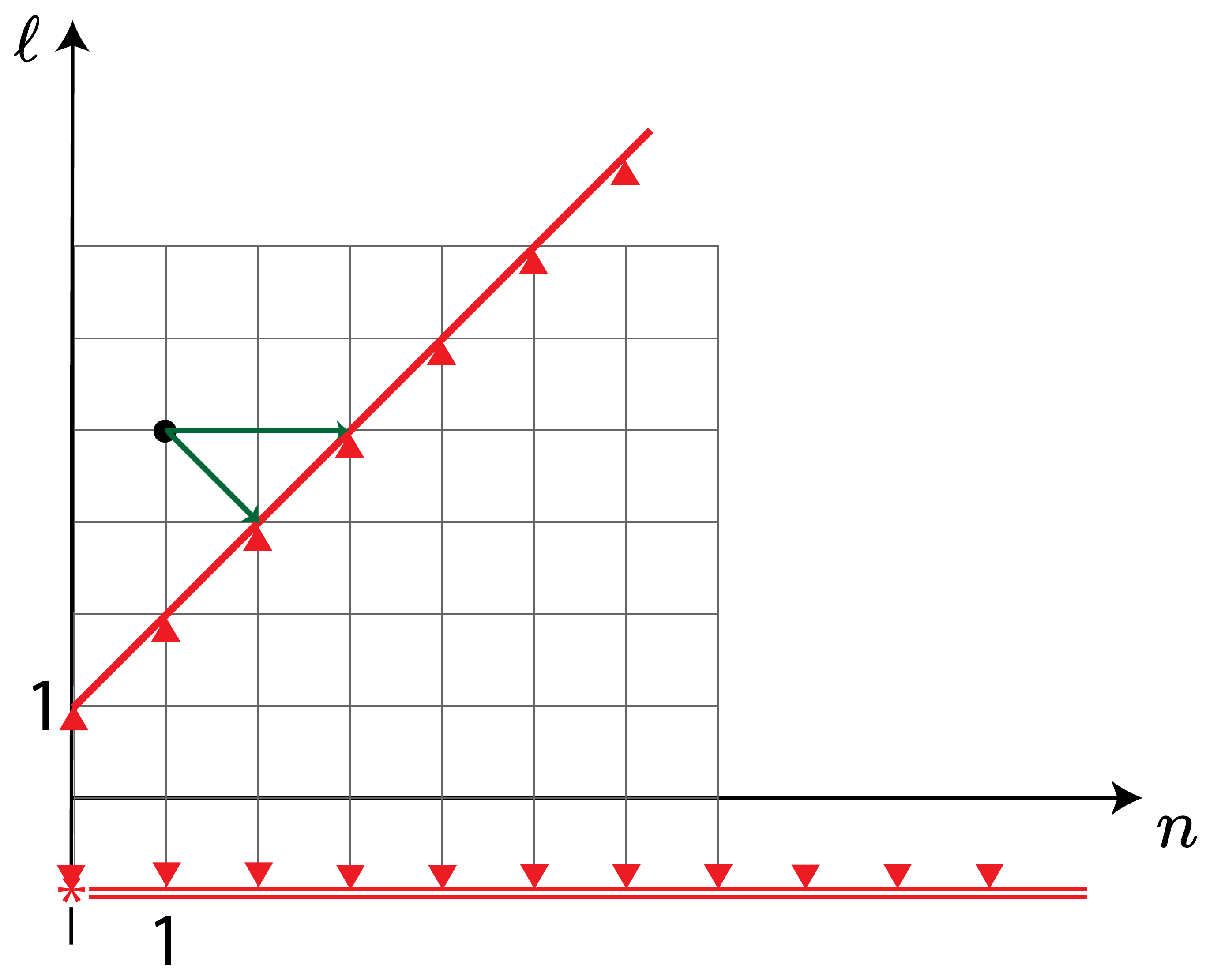}
	\caption{Investigation of our algorithm for the indefinite integral $\int z^4 \calD_\mu(z) \calbD_{\mu+1}(z) \,d z$. The starting point is marked as a black dot and lies in region I. One application of the recursion relation (creating two strands represented by the green arrows) is necessary so that all paths reach a shoreline.}
	\label{fig:algpos:ex_regI}
\end{figure}

As a preliminary step to establishing our result, let us first look at two examples with $\ell + n$ odd, picking one example for each region. 
\paragraph{Example 1} In region I, we take the example $\ell = 4, n = 1$, so we look e.g. at the indefinite integral 
\[
\int z^4 \calD_\mu(z) \calbD_{\mu+1}(z) \,d z
\]
for any real $\mu$. The necessary steps to reach the shorelines are depicted in Fig. \ref{fig:algpos:ex_regI}.

The explicit calculation follows by using the concrete form of the recursion relation, e.g. for $\calD = \calbD = K$ we get
\begin{align}
	\int z^4 &K_\mu(z) K_{\mu+1}(z) \,d z = \nonumber\\	
	&\int z^4 K_\mu(z) K_{\mu+3}(z) \,d z - 2(\mu+2) \int z^3 K_\mu(z) K_{\mu+2}(z) \,d z.
\end{align}

Now we have reached the shoreline with both strands and can apply eq. \eqref{eq:schaf:spcl_KK} to get
\begin{align}
	\int z^4 &K_\mu(z) K_{\mu+1}(z) \,d z = \nonumber\\
	&\frac{z^5}{8} \Bigl(K_\mu(z) K_{\mu+3}(z) - K_{\mu-1}(z) K_{\mu+4}(z)\Bigr) \nonumber\\
    &\qquad- \frac{\mu+2}{3} z^4 \Bigl(K_\mu(z) K_{\mu+2}(z) - K_{\mu-1}(z) K_{\mu+3}(z)\Bigr),
\end{align}
which is valid for all real $\mu$.

\begin{figure}[b]
	\centering
	\includegraphics[width=0.5\textwidth]{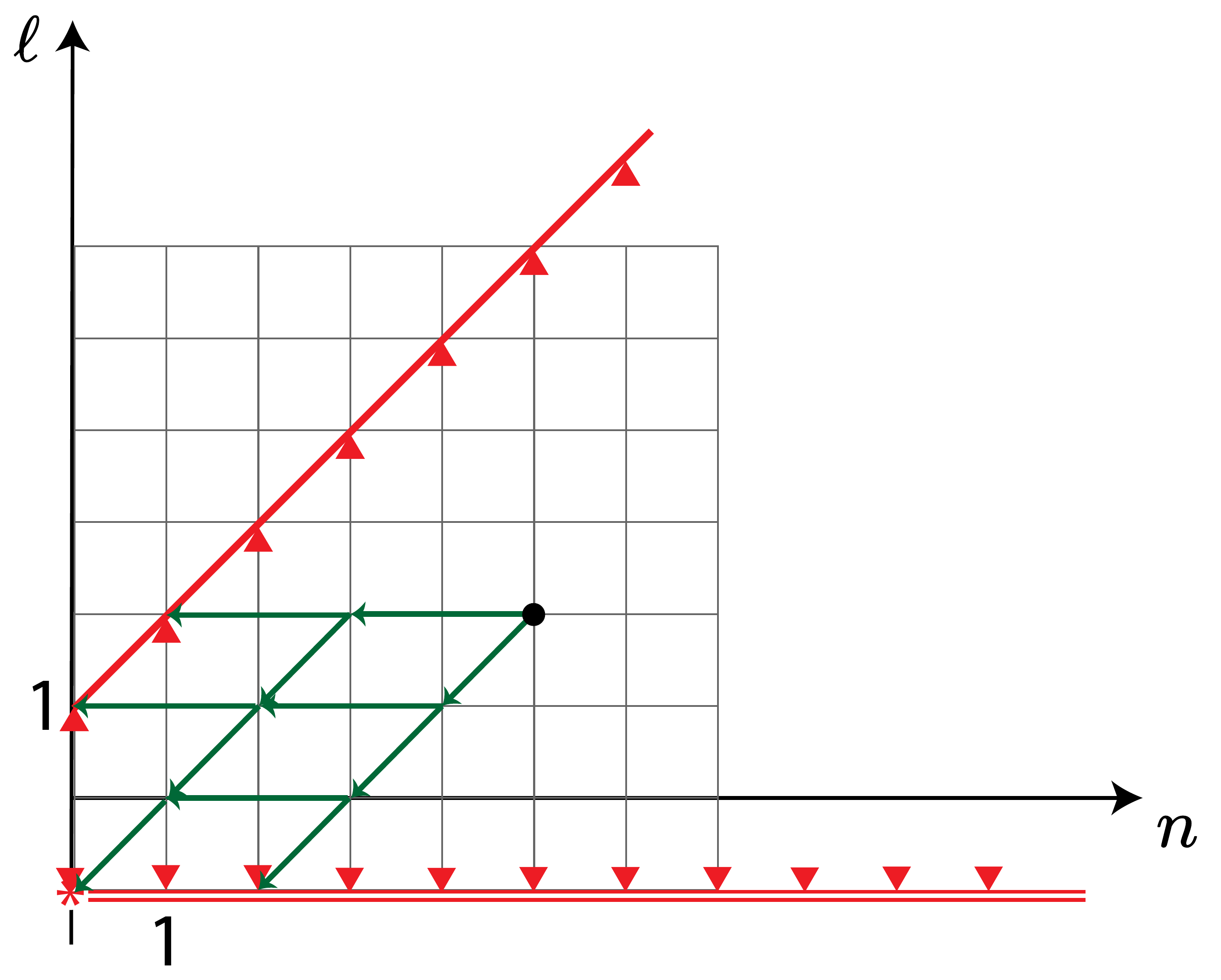}	
	\caption{Investigation of our algorithm for the indefinite integral $\int z^2 \calD_\mu(z) \calbD_{\mu+5}(z) \,d z$. The starting point is marked as a black dot and lies in region II. Application of the recursion relation creates various strands (represented by green arrows) that all terminate on a shoreline. Notice that the integral $\int z^{-1} \calD_\mu(z) \calbD_\mu(z) \,d z$, indicated by the red star, is reached, too.}
	\label{fig:algpos_ex_regII}
\end{figure}

\paragraph{Example 2} As an example for region II consider the integral 
\[
\int z^2 \calD_\mu(z) \calbD_{\mu+5}(z) \,d z.
\]
One possible set of steps for the algorithm is depicted in fig. \ref{fig:algpos_ex_regII}. Here, all strands end on a shoreline, but is has to be noted that the integral $\int z^{-1} \calD_\mu(z) \calbD_\mu(z) \,d z$ is involved. Thus, it depends on the value of $\mu$ whether we get a simple result: for non-zero integer $\mu$ we get a finite amount of terms like the integrand, for non-integer $\mu$ and functions of the first kind we get infinitely many, as by eq. \eqref{eq:lom:ell-1_equal_order}, and for all other cases, no easy form can be given. In that case, one needs other special functions to express the antiderivative, including Hypergeometric and Meijer-Functions.

\paragraph{}We now come to the generalization of these examples. We say an antiderivative is \textit{like the integrand}, if it is a finite sum of terms of the same form as the integrand, that is a product of a monomial and two modified Bessel functions. If the exponent in the monomial of the integrand is non-negative, so is the exponent in all summands. If the antiderivative is an infinite summation of such terms, we say that it possesses the property of being like the integrand \textit{weakly}.

\paragraph{Claim 1} Let us now formulate our first claim: the antiderivative has the property of being like the integrand, in the case that the combination $(n, \ell)$ belongs to the odd sub-lattice, i.e. $n + \ell$ is odd, and is part of region I, i.e.\ $\ell \ge n+1$, or belongs to the odd sub-lattice, is part of region II (i.e.\ $\ell < n+1$) and $\mu$ is a non-zero integer. If in the latter case only functions of the first kind are involved and $\mu$ is non-integer, then the antiderivative has the property of being like the integrand weakly.

For the proof, we will use the well-established concept of the Manhattan metric on the grid of integer $(n, \ell)$-values. We note that each component of all steps has a Manhattan norm of two and thus the parity of the sub-lattice is preserved under all these operations. The shoreline $\ell = n+1$ belongs to the odd sub-lattice. Therefore, it can be reached by any point of the odd sub-lattice in region I where there are no restrictions to the allowed operations. For region II one has to be a bit more careful since one may not move along the grid as freely due to the restrictions of Schafheitlin. As these include that it is not possible to reach the $\ell=n+1$-shoreline from below, one has to use also the regular recursion relation of the Bessel functions. Now again the same argument as before applies, but in addition to reaching the shoreline of $\ell = n + 1$ there are strands that move parallel to this shoreline in region II. These strands obey $\ell = n - (2k+1)$, $k = 0, \dots$, where there will always be the $k=0$-strand when starting from region II. This strand includes the point $(n, \ell) = (1, 0)$, which by eq. \eqref{eq:algpos:rec0_II} can be recursed down to the point $(n, \ell) = (0, -1)$. The corresponding antiderivative is like the integrand in the case of non-zero $\mu$ and has this property weakly in the case of non-integer $\mu$ and functions of the first kind according to eqs. \ref{eq:lom:ell-1_equal_int_order_II} and following. All other parallel strands terminate at the $\ell=-1$-shoreline and give antiderivatives like the integrand.

Consider now points of the even sub-lattice. First, it is obvious that one can reach region II from any point of the even sub-lattice in region I, simply by making repeated use of the regular recursion relation for the modified Bessel functions. This will create numerous strands, which either terminate on a point of the even sub-lattice in region II or end on the $\ell = n+1$-shoreline. Hence, let us assume without loss of generality that we start from a point $(n, \ell)$ of the even sub-lattice in region II.

\begin{figure}[t]
	\centering
	\includegraphics[width=0.5\textwidth]{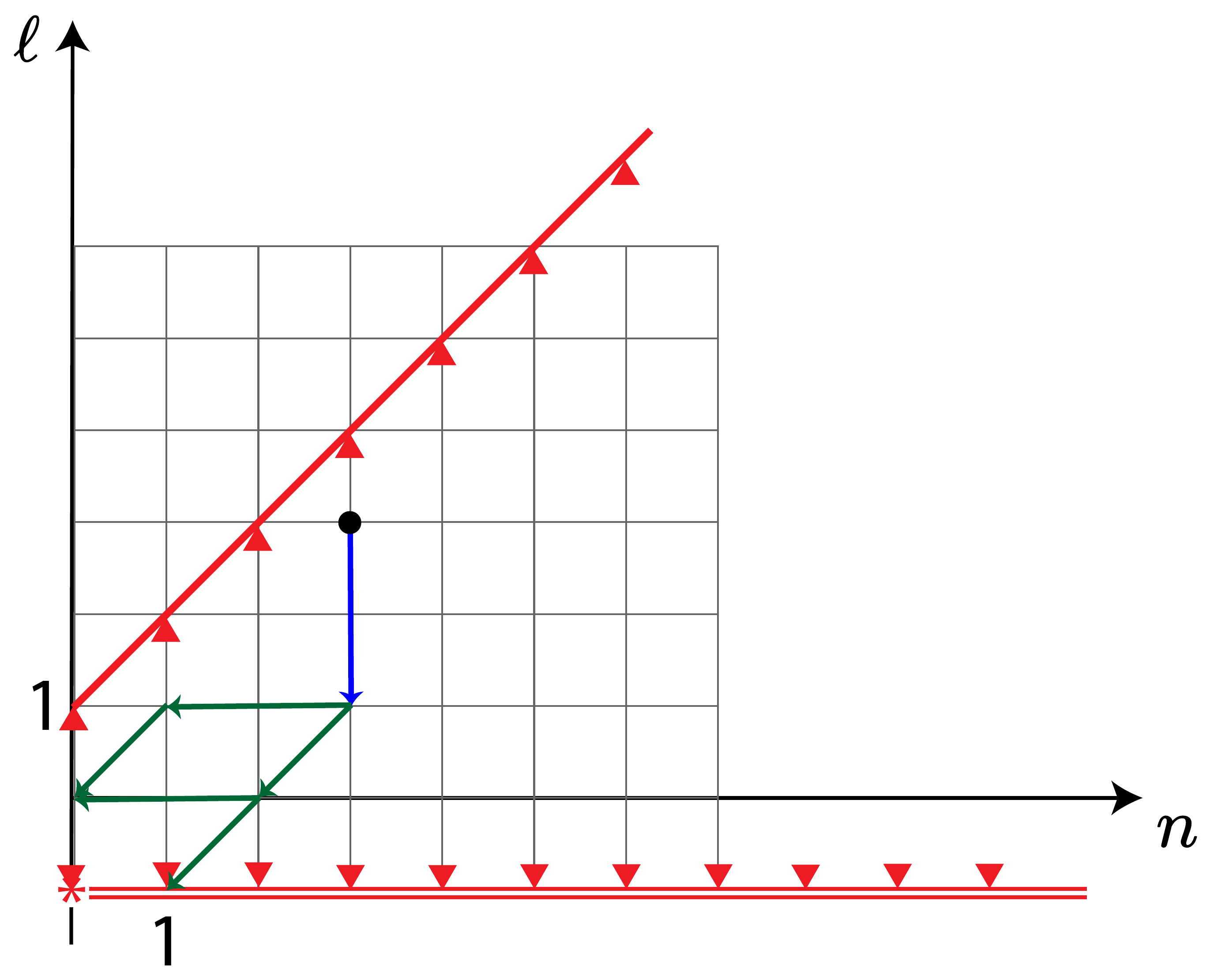}	
	\caption{Investigation of our algorithm for the indefinite integral $\int z^3 \calD_\mu(z) \calbD_{\mu+3}(z) \,d z$. The starting point is marked as a black dot and lies in region II. Application of Schafheitlin (represented by blue arrow) first and then the recursion relation creates various strands (represented by green arrows) that all terminate either on a shoreline or in the origin.}
	\label{fig:algpos:ex_regII_even}
\end{figure}

\paragraph{Example 3} We again start with a concrete example, illustrated in Fig.\ \ref{fig:algpos:ex_regII_even}. One first uses Schafheitlin to get to indefinite integrals with $\ell = 1$. Then, one uses the recursion relation several times, where all strands terminate either on the $\ell=-1$-shoreline or in the origin. This procedure leads us to the claim that for points of the even sub-lattice, i.e. $\ell + n$ even, one can always reduce the antiderivative to terms like the integrand plus the integral represented by the origin in the $\ell$-$n$-plane, namely $\int \calD_\mu(z) \calbD_\mu(z) \,d z$. In formulae, we first have used our extended version of Schafheitlin's reduction formula, eq. \eqref{eq:schafheitlin_full}, which for the case of $\mu = 7/4$ and two modified Bessel functions of the second kind reads
\begin{align}
	\int z^3 K_{7/4}(z) K_{19/4}(z) \,d z = \frac{1}{12} &\Biggl\{\frac{807}{8} \int z K_{7/4}(z) K_{19/4}(z) \,d z + 2 z^3 \deriv{}{z} K_{7/4}(z) K_{19/4}(z) \nonumber\\
	&\qquad+ \frac{173}{8} z^2 K_{7/4}(z) K_{19/4}(z) \nonumber\\
    &\qquad- 2 z^4 \Bigl(K_{7/4}'(z) K_{19/4}'(z) - K_{7/4}(z) K_{19/4}(z)\Bigr) \nonumber\\
	&\qquad+ \frac{3}{4} z^3 \Bigl(K_{7/4}(z) K_{19/4}'(z) - K_{7/4}'(z) K_{19/4}(z)\Bigr)\Biggr\}.
\end{align}

One may now iterate the recursion relations to get for the remaining integral
\begin{align}
	\int z K_{7/4}(z) K_{19/4}(z) \,d z &= \frac{9}{4} \int K_	{7/4}^2(z) \,d z - \frac{z}{2} K_{7/4}^2(z) \nonumber\\
	&\qquad+ \frac{15}{2} \Biggl\{\int K_{7/4}^2(z) \,d z + \frac{11}{2} \int z^{-1} K_{7/4}(z) K_{11/4}(z) \,d z\Biggr\}. 
\end{align}
With an application of Lommel~1 we reach
\begin{align}
	\int z K_{7/4}(z) K_{19/4}(z) \,d z &= \frac{39}{4} \int K_{7/4}^2(z) \,d z - \frac{z}{2} K_{7/4}^2(z) \nonumber\\
	&\qquad- \frac{55}{6} z \Bigl(K_{7/4}'(z) K_{11/4}(z) - K_{7/4}(z) K_{11/4}'(z)\Bigr),
\end{align}
so that the only remaining antiderivative is represented by the origin. One may finally use the recursion relation of Lommel~2 to transfer this into an integral of the form $\int \calD_\lambda(z) \calbD_\lambda(z) \,d z$ with $0 \le \lambda < 1$. In our example,
\begin{align}
	\int z K_{7/4}(z) K_{19/4}(z) \,d z &= -\frac{39}{4} \int K_{3/4}^2(z) \,d z - \frac{22}{5} z K_{7/4}^2(z) + \frac{39}{10} z K_{3/4}^2(z) \nonumber\\
	&\qquad- \frac{55}{6} z \Bigl(K_{7/4}'(z) K_{11/4}(z) - K_{7/4}(z) K_{11/4}'(z)\Bigr),
\end{align}
This last integral $\int \calD_\lambda(z) \calbD_\lambda(z) \,d z$ with $0 \le \lambda < 1$ can be given by special functions. In the case $\lambda = 0$ it generally includes a Meijer-function, which, in the case that only Bessel functions of the first kind are involved, may be simplified to a Hypergeometric function. In the special case $\lambda = 1/2$ it involves the exponential integral. In the other cases, it can be given as a sum where each term involves one Hypergeometric function.

\paragraph{Claim 2}The example above motivates our second important claim, that for points of the even sub-lattice, i.e. $\ell + n$ is even, one can always reduce the antiderivative to terms like the integrand plus the integral represented by the origin in the $\ell$-$n$-plane, namely $\int \calD_\mu(z) \calbD_\mu(z) \,d z$.

To prove this claim, we start from a point of the even sub-lattice in region II. By repeated application of Schafheitlin, one can always reach either the horizontal line at $\ell = 1$ or at $\ell = 0$. In the case of $\ell = 0$, $n$ has to be even, and thus repeated use of the regular recursion relation for the modified Bessel functions creates a strand that we can terminate in the origin. All other strands created terminate on the $\ell = -1$-shoreline, and since $n$ is always even, the point indicated by the star is never reached. Hence, all these integrals are known by Lommel~1, and the only remaining antiderivative is that represented by the origin.
	
The case $\ell = 1$ implies that $n$ is odd. Applying the regular recursion relation for the modified Bessel function once obtains two strands. One goes to $n-2$ at $\ell = 1$, the other goes to $\ell = 0$ with even $n$, and thus for the latter strand the argument from above can be reused. For the former strand, we can again use the regular recursion relation, and get again two strands, one that lands on the $\ell = 0$ horizontal line and can be treated as above. Iterating this procedure, one gets several strands on the $\ell = 0$ horizontal line (which we know from above can lead to the origin), and we terminate one strand at $(n, \ell) = (1, 1)$. From there, we can use the special form of the recursion to go to the origin with only one strand. Therefore, also here, all integrals are known but for the antiderivative represented by the origin. This concludes the proof.

\subsection{A closer look at the generalized Schafheitlin recursion relation\label{sec:shaf_reloaded}}

Before we continue with an algorithm for negative powers, let us examine our extension of Schafheitlin's recursion Eq.\ (\ref{eq:schafheitlin_full}) more closely. At the end of section \ref{sec:schafheitlin} we found that the recursion relation collapses in a specific case and this gave us one of the shorelines of the last section. Now, we want to look at this behaviour more systematically. The recursion relation collapses when one of the prefactors of the integrals vanishes for a specific combination of $\rho$, $\mu$ and $\nu$.

The prefactor of the first integral obviously vanishes for $\rho = -2$. The other prefactor then factorizes,
\begin{align}
	\Biggl[\frac{(\mu^2-\nu^2)^2}{\rho+1} &+ (\rho+1) \Bigl((\rho+1)^2 - 2(\mu^2+\nu^2)\Bigr)\Biggr]_{\rho=-2} = \nonumber\\
    &-\Bigl(\mu+\nu+1\Bigr)\Bigl(\mu+\nu-1\Bigr)\Bigl(\mu-\nu+1\Bigr)\Bigl(\mu-\nu-1\Bigr).
\end{align}
One hence gets a formula for integrals of the kind $\int z^{-2} \calD_\mu(z) \calbD_\nu(z) \,d z$, which however becomes trivial ($0 = 0$) when one of the brackets above vanishes, i.e. when $|\mu - \nu| = 1$ or when $|\mu + \nu| = 1$. Then, no antiderivative can be deduced. In all other cases, the formula is given by
\begin{align}
	\Bigl(\mu+\nu+1\Bigr)&\Bigl(\mu+\nu-1\Bigr)\Bigl(\mu-\nu+1\Bigr)\Bigl(\mu-\nu-1\Bigr) \int \frac{\calD_\mu(z) \calbD_\nu(z)}{z^2} d z = \nonumber\\
	&- \deriv{}{z} \Bigl(\calbD_\mu(z) \calD_\nu(z)\Bigr) - \Bigl(1-\mu^2-\nu^2\Bigr) z^{-1} \calbD_\mu(z) \calD_\nu(z) \nonumber\\
	&- 2z\Bigl(\calbD_\mu'(z) \calD_\nu'(z) - \calbD_\mu(z) \calD_\nu(z)\Bigr) \nonumber\\
	&- (\mu^2 - \nu^2) \Bigl(\calbD_\mu(z) \calD_\nu'(z) - \calbD_\mu'(z) \calD_\nu(z)\Bigr).
    \label{eq:1ovz2}
\end{align}

To assess when the prefactor of the second integral, namely 
\begin{equation}
	\Biggl[\frac{(\mu^2-\nu^2)^2	}{\rho+1} + (\rho+1) \Bigl((\rho+1)^2 - 2(\mu^2+\nu^2)\Bigr)\Biggr],	
\end{equation}
vanishes, let us solve the equation $[\dots]=0$. Rewriting this yields
\begin{equation}
	(\mu^2 - \nu^2)^2 = -(\rho+1)^4 + 2 (\rho+1)^2 (\mu^2 + \nu^2).	
\end{equation}
On setting $r = (\rho + 1)^2$ this becomes the quadratic equation
\begin{equation}
	0 = r^2 - 2(\mu^2 + \nu^2) r + (\mu^2 - \nu^2)^2
\end{equation}
with the solutions $r = (\mu \pm \nu)^2$. Hence, $\rho = \pm (\mu \pm \nu) - 1$. When putting $\nu = \mu + n$, since $\ell = \rho + 2$, we get the case of the shoreline for both negative signs. But we also get three other shorelines, two of which directly resemble the special cases where the recursion relation of Lommel~2 collapsed. However, this also leads to further caveats when applying Schafheitlin: one may not reach any of these shorelines from below. This gives rise to an additional $\mu$- and $\nu$-dependence in the following second algorithm.

Before that, let us however point out that using these results, one can obtain many antiderivatives with negative powers by using Schafheitlin to recurse up to either the $z^{-1}$- or the $z^{-2}$-case. In all cases other than the caveats discussed before, one can then find an antiderivative.

\paragraph{Example 4} For instance one may look at the following indefinite integral and apply our extended version of Schafheitlin's integral recursion once to obtain
\begin{align}
	\int \frac{K_{2/3}(z) K_1(z)}{z^3} d z &= \frac{162}{385} \Biggl\{4 \int \frac{K_{2/3}(z) K_1(z)}{z} d z - \frac{2}{z} \deriv{}{z} \Bigl(K_{2/3}(z) K_1(z)\Bigr) \nonumber\\
	&\qquad - \frac{23}{9z^2} K_{2/3}(z) K_1(z) - 2 \Bigl(K_{2/3}'(z) K_1'(z) - 	K_{2/3}(z) K_1(z)\Bigr) \nonumber\\
	&\qquad+ \frac{5}{18z} \Bigl(K_{2/3}(z) K_1'(z) - K_{2/3}'(z) K_1(z)\Bigr)\Biggr\}.
\end{align}
On using the result of Lommel~1 for the remaining antiderivative we obtain finally
\begin{align}
	\int \frac{K_{2/3}(z) K_1(z)}{z^3} d z &= \frac{162}{385} \Biggl\{-\frac{2}{z} \deriv{}{z} \Bigl(K_{2/3}(z) K_1(z)\Bigr) - \frac{23}{9z^2}	 K_{2/3}(z) K_1(z) \nonumber\\
	&\qquad - 2 \Bigl(K_{2/3}'(z) K_1'(z) - K_{2/3}(z) K_1(z)\Bigr) \nonumber\\
	&\qquad + \Bigl(\frac{36z}{5} + \frac{5}{18z}\Bigr) \Bigl(K_{2/3}(z) K_1'(z) - K_{2/3}'(z) K_1(z)\Bigr)\Biggr\}.
\end{align}

 \subsection{An algorithm to find antiderivatives for a class of non-integer powers\label{sec:non_int_pow}}

Next, we investigate the indefinite integrals
\begin{equation}\label{eq:ni:intdef}
    \int \frac{1}{z^{\sigma+\tau-1-\ell}} \calD_\sigma(z) \calbD_\tau(z) \,d z
\end{equation}
for non-negative integer $\ell$ and real $\sigma$ and $\tau$.

\begin{figure}[ht]
	\centering
	\includegraphics[width=0.7\textwidth]{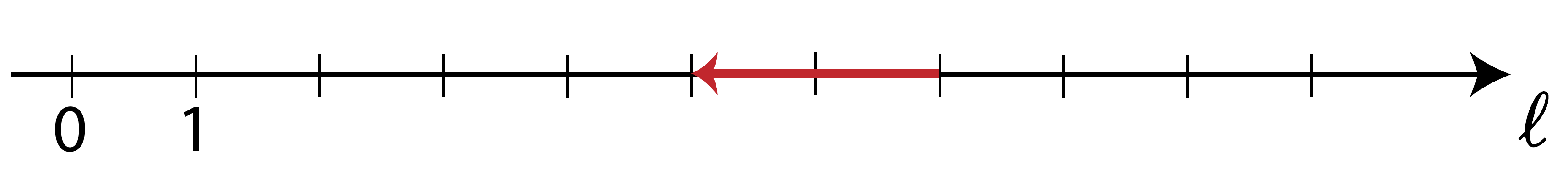}
	\caption{Illustration of the Lommel~2 recursion relation on the $\ell$-ray for our indefinite integrals as defined in eq. \eqref{eq:ni:intdef}.} \label{fig:ni:step}
\end{figure}

We will use the Lommel~2 recursion relations (\ref{eq:lom2recII}--\ref{eq:lom2recKI}) here, which can be brought to the following form by defining $\sigma = \mu + 1$ and $\tau = \nu + 1$ (for concreteness, we give here the case of the mixed product of modified Bessel functions),
\begin{align}
	\ell \int \frac{1}{z^{\mu+\nu-1-(\ell-2)}} K_\mu(z) I_\nu(z) \,d z &- (2\sigma + 2\tau - \ell - 2) \int \frac{1}{z^{\sigma+\tau-1-\ell}} K_\sigma(z) I_\tau(z) \,d z \nonumber\\
	&= \frac{1}{z^{\sigma+\tau-2-\ell}} \Bigl(K_{\sigma-1}(z) I_{\tau-1}(z) + K_{\sigma}(z) I_{\tau}(z)\Bigr). \label{eq:ni:lommel2}
\end{align}
In this way, the added term to the exponent changes by two from $\ell$ to $\ell - 2$. This is illustrated in fig. \ref{fig:ni:step} and shows the only necessary step.

\begin{figure}[ht]
	\centering
	\includegraphics[width=0.7\textwidth]{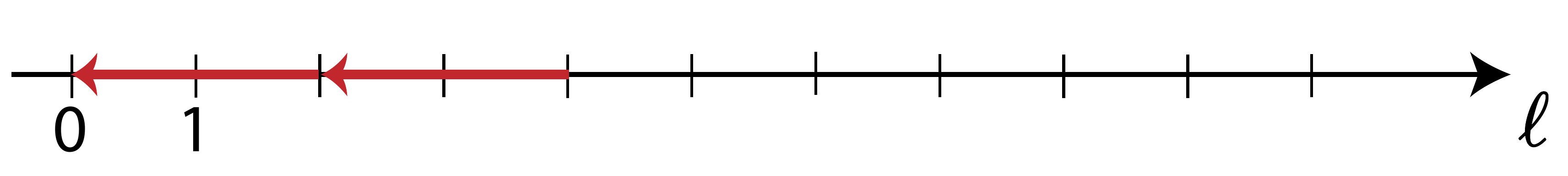}
	\caption{Illustration of the example we calculate in eq. \eqref{eq:ni:ex}. Two applications of the recursion relation in eq. \eqref{eq:ni:lommel2} are shown as red arrows and bring the integral into a form known.}
	\label{fig:ni:ex}
\end{figure}

\paragraph{Example 5} Let us again start by looking at an example,
\begin{equation}\label{eq:ni:ex}
	\int \frac{1}{z^{\sigma+\tau-5}} K_\sigma(z) I_\tau(z) \,d z.
\end{equation}
Here, $\ell = 4$ and we can thus apply the recursion in eq. \eqref{eq:ni:lommel2} twice. This is illustrated in fig. \ref{fig:ni:ex}. The first application yields
\begin{align}
	\int \frac{1}{z^{\sigma+\tau-5}} K_\sigma(z) I_\tau(z) &\,d z = \nonumber\\
    \frac{1}{2(\sigma+\tau-3)} \Biggl\{&4 \int \frac{1}{z^{\mu+\nu-3}} K_\mu(z) I_\nu(z) \,d z \nonumber\\
	&\qquad- \frac{1}{z^{\sigma+\tau-6}} \Bigl(K_{\sigma-1}(z) I_{\tau-1}(z) + K_\sigma(z) I_\tau(z)\Bigr)\Biggr\}
\end{align}
and the second one gives, where $\alpha := \sigma - 2$ and $\beta := \tau - 2$,
\begin{align}
	\int \frac{1}{z^{\mu+\nu-3}} K_\mu(z) I_\nu(z) &\,d z = \nonumber\\
    \frac{1}{2(\sigma+\tau-4)} \Biggl\{&2 \int \frac{1}{z^{\alpha+\beta-1}} K_\alpha(z) I_\beta(z) \,d z \nonumber\\
	&\qquad- \frac{1}{z^{\sigma+\tau-6}} \Bigl(K_{\sigma-2}(z) I_{\tau-2}(z) + K_{\sigma-1}(z) I_{\tau-1}(z)\Bigr)\Biggr\}.
\end{align}
The remaining integral is now calculable by one of our special cases of the last section. It is given by
\begin{equation}
	\int \frac{1}{z^{\alpha+\beta-1}} K_\alpha(z) I_\beta(z) \,d z = \frac{-1}{2(\sigma+\tau-5) z^{\sigma+\tau-6}} \Bigl(K_{\sigma-3}(z) I_{\tau-3}(z) + K_{\sigma-2}(z) I_{\tau-2}(z)\Bigr).
\end{equation}
Hence, on insertion and simplification we finally find
\begin{align}
	\int \frac{1}{z^{\sigma+\tau-5}} K_\sigma(z) I_\tau(z) \,d z =& \nonumber\\
    -\frac{1}{(\sigma+\tau-3) z^{\sigma+\tau-6}} \Biggl\{&\frac{1}{\sigma+\tau-4} \Biggl[\frac{1}{\sigma+\tau-5} K_{\sigma-3}(z) I_{\tau-3}(z) \nonumber\\
	&\qquad + \frac{\sigma+\tau-2}{2} K_{\sigma-1}(z) I_{\tau-1}(z)\Biggr] + \nonumber\\
	&\frac{1}{\sigma+\tau-5} K_{\sigma-2}(z) I_{\tau-2}(z) + \frac{1}{2} K_\sigma(z) K_\tau(z)\Biggr\}.
\end{align}
One can see that this formula is no longer valid when $(\sigma + \tau)\in\{3, 4, 5\}$. We will address this issue more generally below.

\paragraph{Claim 3} Now let us establish our main result of this section, namely that the antiderivative in eq. \eqref{eq:ni:intdef} is like the integrand if $\ell$ is even, and $\sigma + \tau \notin \{\frac{\ell}{2} + k \,|\, k = 1, \dots, \frac{\ell}{2} + 1\}$. The reason is simple: the recursion relation in Eq.\ \eqref{eq:ni:lommel2} (valid under the given prerequisites) always reduces $\ell$ by two units, and the case $\ell = 0$ is known by the special case of Lommel 2 (see Eq.\ (\ref{eq:lom2:specb})) and the antiderivative is like the integrand. Thus, if $\ell$ is divisible by two or in other words an even number, then the indefinite integral is explicitly expressible and all terms of the antiderivative are like the integrand.

It shall be noted that there are some cases where $\sigma + \tau \in \{\frac{\ell}{2} + k \,|\, k = 1, \dots, \frac{\ell}{2} + 1\}$ and the antiderivative is still like the integrand. This is especially the case for integer or half-integer values of both $\sigma$ and $\tau$. Notice that the resulting exponent in the integrands in the denominator, $\sigma + \tau - 1 + \ell$ is always a non-positive integer in these cases, and $\sigma - \tau$ is also an integer. Thus, we are in the case of section \ref{sec:algpos} and, on top of that, since $\ell$ is even, the sum of the exponent and orders of Bessel functions is always odd. From the result of the quoted section, the antiderivative in these cases is like the integrand under the given additional constraints of the quoted section.

\subsection{Summary of our findings on antiderivatives}
In this section we were able to use the method of recursion relations to compute several general types of antiderivatives involving products of two modified Bessel functions. Apart from the well-known recursion relations of the Bessel functions themselves, we could extend a recursion relation originally due to Schafheitlin and thus state a new and far more general integral recursion relation; see Eqs.\ (\ref{eq:schafheitlin_full}--\ref{eq:schafheitlin_fullB}). As an additional tool, we proved the set of recursion relations Eqs.(\ref{eq:algpos:rec0_II}, \ref{eq:algpos:rec0_KK}--\ref{eq:algpos:rec0_IK}). 

Starting from results that were derived by Lommel more than a hundred years ago, we investigated several classes of indefinite integrals using our recursive method. This led us to discover the closure property that, under a condition derived here on the parity of the sum of the degree of a monomial and the orders of the two Bessel functions, the antiderivative can be written as a finite sum of terms of the same form as the integrand. For example, for a monomial with non-negative integral exponent and two modified Bessel functions of the first kind and integer order, this is possible if the parity of the sum of the degree of the monomial and the two orders of Bessel functions is odd. Moreover, in the other cases our approach still allows for a systematic way to simplify other antiderivatives to a common master integral, such as $\int I_0^2(z) \, dz$, that is to our understanding only expressible in terms of more general special functions. On the way, we have given many practical examples of finding antiderivatives, including cases which are not readily found in tables or using a computer algebra system. In particular, Eq.\ (\ref{eq:1ovz2}) is, to our knowledge, a new result.

After this in-depth inspection of a class of indefinite integrals, we can now apply the obtained results to those cases that arise in coordinate-space perturbation theory.

\section{The three-point function in coordinate space \label{sec:3ptFct}}

As a first illustration,
our main aim in this section is to calculate, to leading order in the coupling $g$, the coefficients in the Gegenbauer polynomial expansion of the three-point function introduced in Eq.\ (\ref{eq:3ptdef}) and illustrated in Fig.\ \ref{fig:3ptFct}, 
\begin{equation}\label{eq:Wwn}
W(x,y) \equiv \int_z G_m^{(\lambda)}(z)\, G_{0}^{(\lambda)}(y-z) \,G_m^{(\lambda)}(x-z) =  \sum_{n=0}^\infty C_n^{(\lambda)}(\hat x\cdot \hat y) \,w_n(|x|,|y|).
\end{equation}
Employing the Gegenbauer expansion (\ref{eq:elements:gamma_n^m}) of two propagators and the orthogonality property (\ref{eq:CnOrthog}) of the polynomials, one arrives at
\begin{equation}
	w_n(|x|,|y|) = \frac{m^\lambda}{2^\lambda \Gamma(\lambda)} \frac{1}{\lambda+n} \int_0^\infty d|z|\, |z|^{\lambda+1} K_\lambda(m|z|) \gamma_n^{(m,\lambda)}(|x|,|z|) \gamma_n^{(0,\lambda)}(|y|,|z|).
\end{equation}
It is convenient at this point to define
\ba \label{eq:I1def}
{\cal I}_1(\lambda,n,z_0) &=& 
\int_0^{z_0} dz  \,z^{n+1} \,K_\lambda(z)\,  I_{\lambda+n}(z) ,
\\ \label{eq:I2def}
{\cal I}_2(\lambda,n,z_0,z_1) &=& \int_{z_0}^{z_1} dz \, \frac{1}{z^{2\lambda+n-1}}  \,K_\lambda(z)\,  I_{\lambda+n}(z),
\\ \label{eq:K1def}
{\cal K}_1(\lambda,n,z_0,z_1) &=& \int_{z_0}^{z_1} dz \,z^{n+1} \,K_\lambda(z)\,  K_{\lambda+n}(z),
\\ \label{eq:K2def}
{\cal K}_2(\lambda,n,z_1) &=&\int_{z_1}^\infty dz \, \frac{1}{z^{2\lambda+n-1}}  \,K_\lambda(z)\,  K_{\lambda+n}(z).
\ea
and calculate the corresponding indefinite integrals in the following. The standard substitution $u = m|z|$ gives, in the case $|x| \ge |y|$,
\begin{align}
    w_n(|x|, |y|) = \frac{m^\lambda \Gamma(\lambda)}{4(2\pi^2)^{\lambda+1} |x|^\lambda} \Biggl\{&\frac{K_{\lambda+n}(m|x|)}{|y|^{n+2\lambda} m^{n+2}} \mathcal{I}_1(\lambda, n, m|y|) + \nonumber\\
    &\frac{K_{\lambda+n}(m|x|)}{|y|^{-n} m^{2-2\lambda-n}} \mathcal{I}_2(\lambda, n, m|y|, m|x|) + \nonumber\\
    &\frac{I_{\lambda+n}(m|x|)}{|y|^{-n} m^{2-2\lambda-n}} \mathcal{K}_2(\lambda, n, m|x|)\Biggr\}.
\end{align}
The indefinite integrals are now computed using the special cases of Lommel~2, possibly identifying $K_\lambda(z)$ with $K_{-\lambda}(z)$ first. We find
\begin{align}\label{eq:I1res}
    \mathcal{I}_1&(\lambda, n, m|y|) \nonumber\\
    &\stackrel{(\ref{eq:schaf:spcl_KI})}{=\!=} \frac{u^{n+2}}{2(n+1)} \Bigl(K_\lambda(u) I_{\lambda+n}(u) + K_{\lambda-1}(u) I_{\lambda+n+1}(u)\Bigr)_{u=0}^{m|y|} \nonumber\\
	&= \frac{(m|y|)^{n+2}}{2(n+1)} \Bigl(K_\lambda(m|y|) I_{\lambda+n}(m|y|) + K_{\lambda-1}(m|y|) I_{\lambda+n+1}(m|y|)\Bigr),
\end{align}
\begin{align}\label{eq:I2res}
    \mathcal{I}_2&(\lambda, n, m|y|, m|x|) \nonumber\\
    &\stackrel{(\ref{eq:lom2:specb})}{=\!=} -\frac{u^{2-2\lambda-n}}{2(2\lambda+n-1)} \Bigl(K_{\lambda-1}(u) I_{\lambda+n-1}(u) + K_\lambda(u) I_{\lambda+n}(u)\Bigr)_{u=m|y|}^{m|x|} \nonumber\\
	&= \, -\frac{m^{2-2\lambda-n}}{2(2\lambda+n-1)} \nonumber\\
	&\qquad \cdot \Biggl\{|x|^{2-2\lambda-n} \Bigl(K_{\lambda-1}(m|x|) I_{\lambda+n-1}(m|x|) + K_\lambda(m|x|) I_{\lambda+n}(m|x|)\Bigr) \nonumber\\
	&\qquad - |y|^{2-2\lambda-n} \Bigl(K_{\lambda-1}(m|y|) I_{\lambda+n-1}(m|y|) + K_\lambda(m|y|) I_{\lambda+n}(m|y|)\Bigr)\Biggr\},
\end{align}
and finally (setting $\rho=-(\mu+\nu)$, and then $\mu=\lambda-1$ and $\nu=\lambda+n-1$ in Eq.\ (\ref{eq:lom2recKK}))
\begin{align}
    \mathcal{K}_2&(\lambda, n, m|x|) \nonumber\\
	&= \frac{u^{2-2\lambda-n}}{2(2\lambda+n-1)} \Bigl(K_{\lambda-1}(u) K_{\lambda+n-1}(u) - K_\lambda(u) K_{\lambda+n}(u)\Bigr)_{u=m|x|}^\infty \nonumber\\
	&= -\frac{(m|x|)^{2-2\lambda-n}}{2(2\lambda+n-1)} \Bigl(K_{\lambda-1}(m|x|) K_{\lambda+n-1}(m|x|) - K_\lambda(m|x|) K_{\lambda+n}(m|x|)\Bigr).
    \label{eq:K2res}
\end{align}
The integration bounds $0$ and $\infty$ do not contribute due to the limiting forms of the Bessel functions for small, respectively large argument (see~\cite{nist}).

For the case $|x| < |y|$, we obtain, after making the standard substitution,
\begin{align}
	w_n(|x|,|y|) = \frac{m^\lambda \Gamma(\lambda)}{4(2\pi^2)^{\lambda+1} |x|^\lambda} \Bigg\{&\frac{K_{\lambda+n}(m|x|)}{|y|^{n+2\lambda} m^{n+2}} \mathcal{I}_1(\lambda, n, m|x|) + \nonumber\\
	&\frac{I_{\lambda+n}(m|x|)}{|y|^{n+2\lambda} m^{n+2}} \mathcal{K}_1(\lambda, n, m|x|, m|y|) + \nonumber\\
	&\frac{I_{\lambda+n}(m|x|)}{|y|^{-n} m^{2-2\lambda-n}} \mathcal{K}_2(\lambda, n, m|y|)\Biggr\}.
\end{align}
The remaining indefinite integral can be calculated similarly to the others,
\begin{align}\label{eq:solnCalK1}
& {\cal K}_1(\lambda,n,m|x|,m|y|) \stackrel{(\ref{eq:schaf:spcl_KK})}{=\!=}
\frac{1}{2(n+1)} 
\\ & \quad\times\Big\{ (m|y|)^{n+2}\Big(K_{\lambda}(m|y|)K_{\lambda+n}(m|y|)- K_{\lambda-1}(m|y|)K_{\lambda+n+1}(m|y|)  \Big)
\nonumber\\ &\quad - (m|x|)^{n+2}\Big(K_{\lambda}(m|x|)K_{\lambda+n}(m|x|)- K_{\lambda-1}(m|x|)K_{\lambda+n+1}(m|x|)  \Big)\Big\}.
\nonumber
\end{align}

After simplifications, we arrive at our final expression,
\begin{eqnarray}
w_n(|x|,|y|) &=& \frac{m^{\lambda-1} \Gamma(\lambda)}{8(2\pi^2)^{\lambda+1}|x|^\lambda}
\Big\{ \theta(|x|-|y|) \frac{1}{ (2 \lambda+n-1)|x|^{2 \lambda +n}} 
\nonumber \\ && \Big[  \frac{2(\lambda+n)|x|^{2\lambda+n}}{ (n+1)|y|^{2 \lambda-1}} K_{\lambda+n}(m |x|) \Big(m |y| \, K_{\lambda-1}(m |y|)  I_{\lambda+n-1}(m |y|)
\nonumber\\ &&    +I_{\lambda+n}(m |y|) [(-2 \lambda-n+1) K_{\lambda-1}(m |y|)+m |y| K_\lambda(m |y|)]\Big)
\nonumber\\ &&   - |x|  K_{\lambda-1}(m |x|) |y|^n  \Big]
\label{eq:w_nFINAL}
\\ && + \theta(|y|-|x|)  \frac{1}{  (n+1)|y|^{2 \lambda}} \Big[\frac{|x|^{n+1}}{ |y|^{n}} K_{\lambda-1}(m |x|)
\nonumber   \\ && + \frac{2 |y| (\lambda+n)}{2 \lambda+n-1}  I_{\lambda+n}(m |x|) \Big(m |y| K_\lambda(m |y|) K_{\lambda+n}(m |y|)
\nonumber     \\ && -K_{\lambda-1}(m |y|) [m |y| K_{\lambda+n-1}(m |y|)+(2 \lambda+n-1) K_{\lambda+n}(m y)]\Big)  \Big] \Big\}.
\nonumber 
\end{eqnarray}
It is expressed in terms of products of two modified Bessel functions, which stems from the diagram involving two massive propagators. In the next two subsections, we show that this result can be reused straightforwardly in physically relevant theories.

It is also worth noting that the three-point function obeys the discrete symmetry relation
\begin{align}
    W(x,y) = W(x,x-y).
\end{align}
The convergence rate of the series in Gegenbauer polynomials depends on the geometry of the triangle formed by the origin and the two points $x$ and $y$.
For an efficient numerical evaluation, it is thus preferable to expand in $C_n^{(\lambda)}(\hat x\cdot \hat y)$ or in $C_n^{(\lambda)}(\hat x\cdot \widehat{x-y})$ depending on which of  $|\log(|x|/|y|)|$ and $|\log(|x|/|x-y|)|$ is larger~\cite{Asmussen:2022oql}. 

\subsection{Three-point function for scalar QED}

Following our study of the three-point function in a scalar field theory, let
\begin{equation}
    \< \phi(x) A_\nu(y) \phi^*(w)\> = ie \Bigl(\partial_\nu^{(x)} {W}_{\rm ext}(x,y,w) - \partial_\nu^{(w)} {W}_{\rm ext}(x,y,w)\Bigr)
\end{equation}
be the corresponding three-point function in scalar QED in Feynman gauge.
The relevant part of the interaction Lagrangian reads $e j_\mu A_\mu$, where $j_\mu = i \Bigl(\bigl(\partial_\mu \phi^*\bigr) \phi - \phi^* \bigl(\partial_\mu \phi\bigr)\Bigr)$ is the particle current. Here,
\begin{equation}
    {W}_{\rm ext}(x,y,w) = \int_z G_0^{(\lambda)}(y-z) G_m^{(\lambda)}(x-z) G_m^{(\lambda)}(z-w),
\end{equation}
and we first consider an arbitrary point $w$: this allows us to use the chain rule to change the $w$-derivative into an $x$- and $y$-derivative, and then set $w = 0$. The procedure finally yields
\begin{equation}
    \< \phi(x) A_\nu(y) \phi^*(0)\> = ie \Bigl(2 \partial_\nu^{(x)} + \partial_\nu^{(y)}\Bigr) W(x,y).
\end{equation}

\subsection{Three-point function for spinor QED}

Now let 
\begin{equation}
\< \psi(x) A_\nu(y) \bar\psi(0)\> = -ie\,
\int_z G_0^{(\lambda)}(y-z)\,S(x-z) \gamma_\nu S(z)
\end{equation}
be the analogous three-point function in spinor QED, which has the interaction term ${\cal L}_{\rm int} = ie A_\mu \bar\psi\gamma_\mu\psi$.
Again, we have restricted ourselves to Feynman gauge.

Exploiting Eq.\ (\ref{eq:Sm1}), as well as translation invariance similarly to the scalar-QED case, we can express the three-point function in terms of $W$,
\begin{equation}
    \< \psi(x) A_\nu(y) \bar\psi(0)\> 
    = (-ie) (-\partial\!\!\!/^{(x)} + m)\,\gamma_\nu\,
    (-\partial\!\!\!/^{(x)}-\partial\!\!\!/^{(y)} + m)\,
    W(x,y).
\end{equation}

\subsection{A variant of the three-point function\label{sec:Wtilde}}

As a three-point function closely related to $W(x,y)$, we define the function 
\begin{align}\label{eq:Wtilde}
    \widetilde W(x,y) = \int d^dz\; 
    G^{(\lambda-1)}_m(z)
    G^{(\lambda)}_{0}(y-z)\, 
    G^{(\lambda)}_{m}(x-z)\, 
    =  \sum_{n=0}^\infty C_n^{(\lambda)}(\hat x\cdot \hat y) \tilde w_n(|x|,|y|).
\end{align}
It will be useful in section (\ref{sec:finiteT}) to compute the one-loop correction to a scalar propagator at finite temperature. 
Following the same procedure as for $W(x,y)$ we obtain, for $|x|>|y|$,
\begin{align}
   \tilde w_n(|x|, |y|) = &
   \frac{m^{\lambda-2} \Gamma(\lambda)}{2^{\lambda+2}\pi^{2\lambda+1} |x|^\lambda} 
   \Biggl\{\frac{K_{\lambda+n}(m|x|)}{|y|^{n+2\lambda} m^{n+2}} \tilde{\mathcal{I}}_1(\lambda, n, m|y|)  \\
    &+\frac{K_{\lambda+n}(m|x|)}{|y|^{-n} m^{2-2\lambda-n}} \tilde{\mathcal{I}}_2(\lambda, n, m|y|, m|x|) + 
    \frac{I_{\lambda+n}(m|x|)}{|y|^{-n} m^{2-2\lambda-n}} \tilde{\mathcal{K}}_2(\lambda, n, m|x|)\Biggr\},
    \nonumber
\end{align}
where the relevant integrals are given by
\begin{align}
    \tilde {\cal I}_1(\lambda,n,x) &{\equiv} \int_0^x dz\,z^{n+2}K_{\lambda-1}(z) I_{\lambda+n}(z)
 \\ & = {\cal I}_1(\lambda-1,n+1,x) 
\nonumber \\ &\stackrel{(\ref{eq:I1res})}{=\!=} \frac{x^{n+3}}{2 (n+2)} (K_{\lambda -2}(x) I_{n+\lambda +1}(x)+K_{\lambda -1}(x) I_{n+\lambda}(x)),
    \nonumber \\
    \tilde {\cal I}_2(\lambda,n,x,y) &\equiv
  \int_x^y dz\, \frac{1}{z^{2\lambda+n-2}} K_{\lambda-1}(z) I_{\lambda+n}(z)
  \\ &= {\cal I}_2(\lambda-1,n+1,x,y)
 \nonumber \\ &\stackrel{(\ref{eq:I2res})}{=\!=} \frac{1}{2(2\lambda+n-2)}
  \nonumber  \\ & \times \bigg\{x^{-2 \lambda -n+3} (K_{\lambda -2}(x) I_{n+\lambda -1}(x)+K_{\lambda -1}(x) I_{n+\lambda
   }(x))
\nonumber   \\ & -y^{-2 \lambda -n+3} (K_{\lambda -2}(y) I_{n+\lambda -1}(y)+K_{\lambda -1}(y)
   I_{n+\lambda }(y))\bigg\},
   \nonumber
   \\ \tilde {\cal K}_1(\lambda,n,x,y) &\equiv \int_x^y dz z^{n+2} \,K_{\lambda-1}(z) \,K_{\lambda+n}(z)
   \\ & = {\cal K}_1(\lambda-1,n+1,x,y) 
\nonumber 
   \\ & \stackrel{(\ref{eq:solnCalK1})}{=\!=} \frac{1}{2(n+2)}
 \nonumber  \\ & \times \bigg\{
   x^{n+3} (K_{\lambda -2}(x) K_{n+\lambda +1}(x)-K_{\lambda -1}(x) K_{n+\lambda
   }(x))
 \nonumber  \\ & +y^{n+3} (K_{\lambda -1}(y) K_{n+\lambda }(y)-K_{\lambda -2}(y) K_{n+\lambda
   +1}(y))
   \bigg\},
   \nonumber
   \\ \tilde {\cal K}_2(\lambda,n,x) &\equiv 
   \int_{x}^\infty dz \frac{1}{z^{2\lambda+n-2}}\,K_{\lambda-1}(z)\,K_{\lambda+n}(z)
  \\ &= {\cal K}_2(\lambda-1,n+1,x)
  \nonumber 
  \\ & \stackrel{(\ref{eq:K2res})}{=} 
   \frac{x^{-2 \lambda -n+3}}{2 (2 \lambda +n-2)} (K_{\lambda -1}(x) K_{n+\lambda }(x)-K_{\lambda -2}(x)
   K_{n+\lambda -1}(x)).
   \nonumber
\end{align}
In the other case, $|x|<|y|$, we arrive at 
\begin{align}
	\tilde w_n(|x|,|y|) = \frac{m^{\lambda-2}\Gamma(\lambda) }{2^{\lambda+2}\pi^{2\lambda+1} |x|^\lambda} \Bigg\{&\frac{K_{\lambda+n}(m|x|)}{|y|^{n+2\lambda} m^{n+2}} \tilde{\mathcal{I}}_1(\lambda, n, m|x|) + \nonumber\\
	&\frac{I_{\lambda+n}(m|x|)}{|y|^{n+2\lambda} m^{n+2}} \tilde{\mathcal{K}}_1(\lambda, n, m|x|, m|y|) + \nonumber\\
	&\frac{I_{\lambda+n}(m|x|)}{|y|^{-n} m^{2-2\lambda-n}} \tilde{\mathcal{K}}_2(\lambda, n, m|y|)\Biggr\}.
\end{align}
After simplifications, our final result reads
\begin{align}\label{eq:wntilde_final}
&\tilde w_n(|x|,|y|) = \frac{m^{\lambda-1}\Gamma(\lambda) }{2^{\lambda+2}\pi^{2\lambda+1} |x|^\lambda}
\\ & \bigg\{\theta(|x|-|y|) \bigg[\frac{ |y|^{-2 \lambda +3}}{2 (n+2)} K_{n+\lambda }(m |x|) (K_{\lambda -2}(m |y|)
   I_{n+\lambda +1}(m |y|)+K_{\lambda -1}(m |y|) I_{n+\lambda }(m |y|))
\nonumber   \\ & +\frac{ |y|^n
    K_{n+\lambda }(m |x|)}{2 (2 \lambda +n-2)} \big( |x|^{-2 \lambda -n+3} (K_{\lambda -2}(m|x|) I_{n+\lambda -1}(m |x|)+K_{\lambda -1}(m |x|) I_{n+\lambda }(m |x|))
\nonumber   \\ & - |y|^{-2 \lambda
   -n+3} (K_{\lambda -2}(m |y|) I_{n+\lambda -1}(m |y|)+K_{\lambda -1}(m |y|) I_{n+\lambda }(m
   |y|))\big)
 \nonumber  \\ & +\frac{|y|^n \, |x|^{-2 \lambda -n+3}}{2 (2 \lambda +n-2)}
   I_{n+\lambda }(m |x|) (K_{\lambda -1}(m |x|) K_{n+\lambda }(m |x|)-K_{\lambda -2}(m |x|)
   K_{n+\lambda -1}(m |x|))\bigg]
\nonumber\\ & +
    \theta(|y|-|x|)\frac{1}{2}  |y|^{-2\lambda-n}\bigg[  \frac{1}{n+2} \Big(m^{-1}|x|^{n+2} K_{\lambda
   -2}(m |x|)
\nonumber   \\ & +  |y|^{n+3} I_{n+\lambda }(m |x|) (K_{\lambda -1}(m |y|) K_{n+\lambda }(m
   |y|)-K_{\lambda -2}(m |y|) K_{n+\lambda +1}(m |y|))\Big)
\nonumber   \\ & -\frac{|y|^{n+3}}{2 \lambda +n-2} I_{n+\lambda }(m |x|) (K_{\lambda -2}(m |y|)
   K_{n+\lambda -1}(m |y|)-K_{\lambda -1}(m |y|) K_{n+\lambda }(m |y|))\bigg]\bigg\}.
   \nonumber
\end{align}

\section{The two-point function at one loop order\label{sec:2pt1loop}}

\begin{figure}[t]
    \centering
    \includegraphics[width=0.36\linewidth]{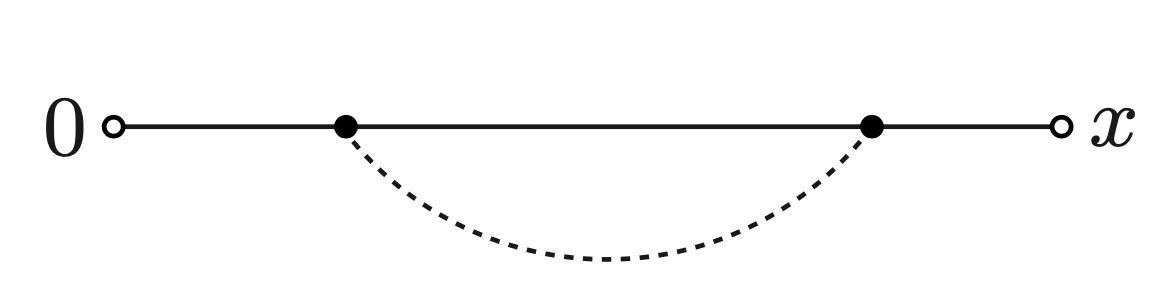}~~~~~~~~~~~~~~
    \includegraphics[width=0.36\linewidth]{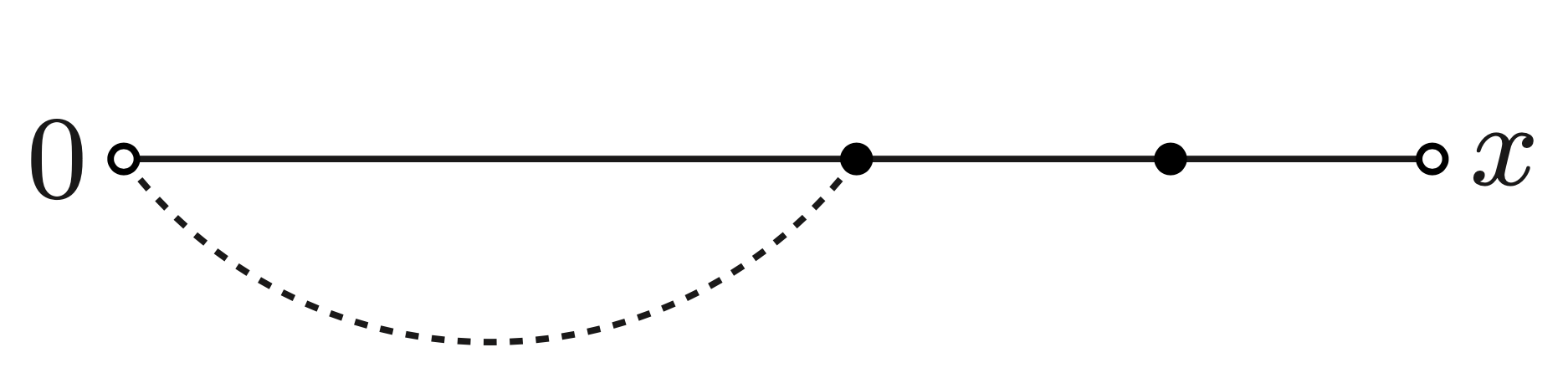}
    \caption{Left: Feynman diagram representing the one-loop contribution to the two-point function in coordinate space and corresponding to expression (\ref{eq:2pt1loop_f1}). Right: an equivalent diagram that is simpler to evaluate and corresponds to expression (\ref{eq:2pt1loop_f2}).}
    \label{fig:2pt1loop}
\end{figure}

As a second illustration of the methodology presented in sections \ref{sec:elements} and \ref{sec:antideriv}, we present a calculation of a two-point position-space function at one-loop order.
We note that the renormalized one-loop position-space electron propagator in QED was obtained in~\cite{ParrinoMasterArbeit} for $\lambda=1$, while a result for general $\lambda$ was derived in~\cite{SchroderMasterArbeit} for a one-loop two-point function in scalar field theory.
In the field theory alluded to in the introduction, consisting of a massive complex scalar field $\phi$ and a massless real scalar field $\chi$ interacting via ${\cal L}_{\rm int}=g\chi \phi^*\phi$, the two-point function can be computed perturbatively according to 
\begin{align}\label{eq:d2GmDef}
    \<\phi(x)\phi^*(0)\> = &\; G_m^{(\lambda)}(x) + 
    g^2 \delta_2G_m^{(\lambda)}(x)  + {\rm O}(g^4).
\end{align}
The leading correction can be evaluated via the following steps, which are illustrated by Fig.\ \ref{fig:2pt1loop},
\begin{align}
\delta_2 G_m^{(\lambda)}(x)  =&\; \frac{1}{2} \int d^dy \int d^dz \<\phi(x) \chi(y)\phi^*(y)\phi(y)\chi(z)\phi^*(z)\phi(z) \phi^*(0)\>_0 
\\ \label{eq:2pt1loop_f1}
=& \;\int d^dy \int d^dz\; G_m^{(\lambda)}(x-y) G_0^{(\lambda)}(y-z)G_m^{(\lambda)}(y-z)G_m^{(\lambda)}(z)
\\ \label{eq:2pt1loop_f2}\stackrel{z=y-w}{=\!\!=\!\!=} & \;\int d^dw\; G_0^{(\lambda)}(w)G_m^{(\lambda)}(w)\int d^dy \; G_m^{(\lambda)}(x-y) G_m^{(\lambda)}(y-w) 
\\ 
\stackrel{{\rm Eq.\,}(\ref{eq:GmStarGm})}{=\!\!=\!\!=} & \;\int d^dw\; G_0^{(\lambda)}(w)G_m^{(\lambda)}(w) \left(-\frac{\partial}{\partial m^2}\right) G_m^{(\lambda)}(x-w)
\label{eq:2pt1loop_f3}
\\ =& \frac{1}{4\pi}\int d^dw\; G^{(\lambda)}_0(w) G^{(\lambda)}_m(w)  G^{(\lambda-1)}_m(x-w).
\label{eq:2pt1loop_f4}
\end{align}
From the form (\ref{eq:2pt1loop_f2}), it is easy to see that the function to be computed obeys the inhomogeneous fourth-order differential equation
\begin{align}\label{eq:PDEd2Gm}
    (-\triangle + m^2)^2\; \delta_2 G_m^{(\lambda)}(x) = G_0^{(\lambda)}(x) G_m^{(\lambda)}(x),
\end{align}
where the triangle stands for the $(d=2\lambda+2)$-dimensional Laplacian.

Proceeding with the calculation from expression (\ref{eq:2pt1loop_f4}), one may insert the expression of the first two propagators and use the expansion (\ref{eq:elements:dm2gamma_n^mB}) of the last propagator, 
yielding
\begin{align}
	\delta_2 G_m^{(\lambda)}(x) = \frac{m^{3\lambda-3} \Gamma(\lambda)}{16 (2\pi^2)^{\lambda+1} \lambda} \frac{1}{|x|^{\lambda-1}} \Biggl\{&K_{\lambda-1}(m|x|) \int_0^{m|x|} \mathrm{d}u\, u^{2-2\lambda} K_\lambda(u) I_{\lambda-1}(u) +	\nonumber\\
	&I_{\lambda-1}(m|x|) \int_{m|x|}^\infty \mathrm{d}u\, u^{2-2\lambda} K_\lambda(u) K_{\lambda-1}(u) - \nonumber\\
	&K_{\lambda+1}(m|x|) \int_0^{m|x|} \mathrm{d}u\, u^{2-2\lambda} K_\lambda(u) I_{\lambda+1}(u) - \nonumber\\
	&I_{\lambda+1}(m|x|) \int_{m|x|}^\infty \mathrm{d}u\, u^{2-2\lambda} K_\lambda(u) K_{\lambda+1}(u)\Biggr\}.
    \label{eq:se_conv_general_ints}
\end{align}
In appendix \ref{sec:ldageq1}, we verify that first setting $\lambda=1$ and carrying out a dedicated calculation leads to the same result as evaluating Eq.\ (\ref{eq:se_conv_general_ints}) for $\lambda=1$.

\subsection{Solving the integrals for $\lambda \ne 1$}
Let $\lambda \ne 1$. 
We will now derive the antiderivatives for the four integrals. For the first, we can immediately apply Lommel~2 (see Eq.\  (\ref{eq:lom2:specb})) to obtain
\begin{equation}
	\int \mathrm{d}u\, u^{2-2\lambda} K_\lambda(u) I_{\lambda-1}(u) = -\frac{u^{3-2\lambda}}{2(2\lambda-2)} \Bigl(K_{\lambda-1}(u) I_{\lambda-2}(u) + K_\lambda(u) I_{\lambda-1}(u)\Bigr).
\end{equation}
Making use of the identity $K_\mu(z) I_{\mu-1}(z) + K_{\mu-1}(z) I_\mu(z) = 1/z$ and applying the recursion relation for the modified Bessel function of the second kind, it follows that
\begin{equation}
	\int \mathrm{d}u\, u^{2-2\lambda} K_\lambda(u) I_{\lambda-1}(u) = -\frac{u^{2-2\lambda}}{2} \Biggl\{\frac{1}{2(\lambda-1)} + I_{\lambda-1}(u) K_{\lambda-1}(u)\Biggr\}.
\end{equation}

Similarly, for the second integral it follows by Lommel~2 that
\begin{align}
	\int \mathrm{d}u\, u^{2-2\lambda} K_\lambda(u) K_{\lambda-1}(u) &= -\frac{u^{2-2\lambda}}{2} K_{\lambda-1}^2(u).
\end{align}

For the third integral, one has to first use the recursion relation of Lommel~2 (see Eq.\ (\ref{eq:lom2recKI})),
\begin{align}
	\int \mathrm{d}u\, u	^{2-2\lambda} K_\lambda(u) I_{\lambda+1}(u) = &-\frac{2}{2(1-2\lambda)} \int \mathrm{d}u\, u^{2-2\lambda} K_{\lambda-1}(u) I_\lambda(u) \nonumber\\
	&\qquad + \frac{u^{3-2\lambda}}{2(1-2\lambda)} \Bigl(K_{\lambda-1}(u) I_\lambda(u) + K_\lambda(u) I_{\lambda+1}(u)\Bigr).
\end{align}
The remaining integral may be evaluated via the special case of Lommel~2 (Eq.\ (\ref{eq:lom2:specb})),
\begin{align}
	\int \mathrm{d}u\, u	^{2-2\lambda} K_\lambda(u) I_{\lambda+1}(u) = \frac{u^{3-2\lambda}}{2(1-2\lambda)} \Biggl\{&\frac{1}{2\lambda-2} \Bigl(K_{\lambda-2}(u) I_{\lambda-1}(u) + K_{\lambda-1}(u) I_\lambda(u)\Bigr) \nonumber\\
	&+ \Bigl(K_{\lambda-1}(u) I_\lambda(u) + K_\lambda(u) I_{\lambda+1}(u)\Bigr)\Biggr\}.
\end{align}
By using the identity $K_\mu(z) I_{\mu-1}(z) + K_{\mu-1}(z) I_\mu(z) = 1/z$ twice and using the recursion relation of the Bessel functions, one can simplify this,
\begin{align}
	\int \mathrm{d}u\, u	^{2-2\lambda} K_\lambda(u) I_{\lambda+1}(u) = \frac{u^{2-2\lambda}}{2(1-2\lambda)} \Biggl\{&\frac{2\lambda-1}{2\lambda-2} - 2\lambda K_\lambda(u) I_\lambda(u) - K_{\lambda-1}(u) I_{\lambda-1}(u)\Biggr\}.
\end{align}

Finally, by similar means one can deduce the antiderivative for the fourth integral. The result is
\begin{align}
	\int \mathrm{d}u\, u^{2-2\lambda} K_\lambda(u) K_{\lambda+1}(u) &= \frac{u^{2-2\lambda}}{2(1-2\lambda)} \Bigl\{2\lambda K_\lambda^2(u) - K_{\lambda-1}^2(u)\Bigr\}.
\end{align}
It shall be noted that the calculation can also be performed for $\lambda = 1$ and yields similar results, but with the inclusion of a logarithm, see appendix.

\subsubsection{Evaluation of the definite integrals}\label{dfnt_integrals}

We may now plug in the integral boundaries into the antiderivatives derived above. We start with the simpler cases, obtaining for the second integral
\begin{equation}
	\int_{m|x|}^\infty \mathrm{d}u\, K_\lambda(u) K_{\lambda-1}(u) = \frac{(m|x|)^{2-2\lambda}}{2} K_{\lambda-1}^2(m|x|),
\end{equation}
because the upper boundary does not contribute due to the long-distance behaviour of the modified Bessel function of the second kind, as described in the last section. 
For the fourth integral the upper boundary does not contribute, either, and hence we get
\begin{equation}
	\int_{m|x|}^\infty \mathrm{d}u\, u^{2-2\lambda} K_\lambda(u) K_{\lambda+1}(u) = - \frac{(m|x|)^{2-2\lambda}}{2(1-2\lambda)} \Bigl(2\lambda K_\lambda^2(m|x|) - K_{\lambda-1}^2(m|x|)\Bigr).	
\end{equation}
For the first and third integral, we consider different cases depending on the value of $\lambda$.

\paragraph{The case $\lambda<1$}

For the first integral, one finds
\begin{align}
   &\int_0^{m|x|} \mathrm{d}u\, u^{2-2\lambda} K_\lambda(u) I_{\lambda-1}(u) =
\frac{4^{-\lambda } \Gamma (1-\lambda )}{\Gamma (\lambda )}
\\ & \qquad\qquad -\frac{1}{2} (m |x|)^{2-2 \lambda } \left(\frac{1}{2 (\lambda-1)}+I_{\lambda -1}(m |x|) K_{\lambda -1}(m |x|)\right),
\nonumber
   \end{align}
while for the third integral, we obtain
   \begin{align}
  & \int_0^{m|x|} \mathrm{d}u\, u^{2-2\lambda} K_\lambda(u) I_{\lambda+1}(u) =
       \frac{4^{-\lambda } \Gamma (1-\lambda )}{\Gamma (\lambda )(1-2 \lambda)}
       \\ & \qquad +\frac{(m |x|)^{2-2 \lambda } }{2 (1-2 \lambda )}
       \left(\frac{2 \lambda -1}{2 \lambda -2}-I_{\lambda -1}(m |x|) K_{\lambda -1}(m |x|)-2 \lambda  I_{\lambda }(m |x|) K_{\lambda
   }(m |x|)\right).
   \nonumber
   \end{align}

Initially, we thus obtain a result in the form of a polynomial of degree three in modified Bessel functions.
After simplifications however, the final result reads
\begin{align}\label{eq:d2GmFINALform1}
    &\delta_2 G_m^{(\lambda)}(x) =
    \frac{m^{3 \lambda -4}  }{ 32 (2\pi^2)^{\lambda +1}(2 \lambda -1)|x|^{\lambda } }
    \\ & \quad\times
    \left( (m |x|)^{2-2 \lambda } 
   \Gamma (\lambda -1) K_{\lambda -2}(m |x|)+4^{1-\lambda}  \Gamma (1-\lambda ) (m |x| K_{\lambda -1}(m |x|)+K_{\lambda }(m |x|))\right)
   \nonumber
    \\ &= \frac{1}{4(4\pi)^{\lambda +1}( \lambda -\frac{1}{2})(\lambda-1)}
\label{eq:d2GmFINALform2}
    \\ & \quad\times
    \left((4\pi)^{\lambda-1}  G_0^{(\lambda)}(x)
    G^{(\lambda-2)}_m(x) - m^{2 (\lambda -2)} \Gamma(2-\lambda ) \left(\frac{m^2}{2\pi} G^{(\lambda-1)}_m(x) + G_m^{(\lambda)}(x) \right)\right).
   \nonumber
\end{align}
As the second expression shows, the $x$-dependence can entirely be expressed in terms of propagators. 
We have verified that it obeys the differential equation (\ref{eq:PDEd2Gm}). We suspect the `reason' for the dramatic simplification to be that the two-point function obeys a dispersive representation involving a one-particle pole and a two-particle continuum.

We have also inspected the case $\lambda > 1$ in the appendix \ref{sec:ldageq1}. Infinities arise when evaluating the antiderivatives at the bounds of integration.

\section{A mixed momentum/coordinate-space amplitude
\label{sec:3ptmised}}

\begin{figure}
    \centering
    \includegraphics[width=0.36\linewidth]{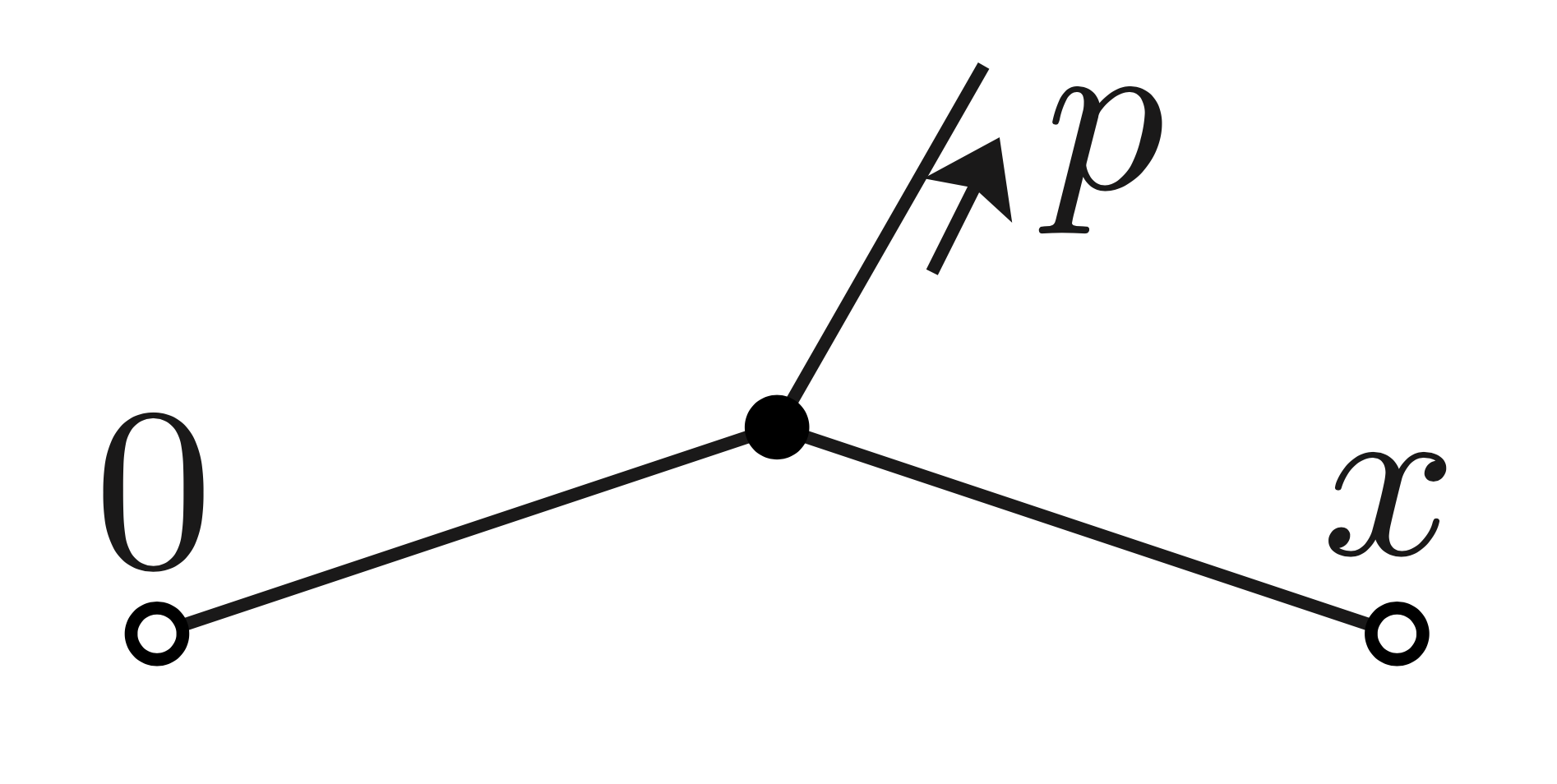}
    \caption{Diagrammatic representation of the mixed coordinate/momentum space three-point function $L^{(\lambda)}_{m,M}(p,x) $ investigated in section~\ref{sec:3ptmised}.}
    \label{fig:3ptmixed}
\end{figure}

In this section, we introduce and compute the function
\be\la{eq:Ldef}
L^{(\lambda)}_{m,M}(p,x) \equiv \int d^dz\, G_m^{(\lambda)}(z) e^{-ipz} G_M^{(\lambda)}(x-z) .
\ee
In terms of its physical interpretation, this function is related to the coordinate-space three-point function computed in section \ref{sec:3ptFct} by a Fourier transform and by `amputating' the amplitude by the propagator of the massless particle outgoing as a plane wave: in the equal-mass case $M=m$, we can write
\be
L^{(\lambda)}_{m,m}(p,x) = p^2 \int d^dy\; e^{-ip\cdot y}\, W(x,y).
\ee
The function $L^{(\lambda)}_{m,M}(p,x)$ is a solution to the inhomogeneous differential equation
\begin{equation}
(-\triangle_x + M^2) L^{(\lambda)}_{m,M}(p,x) = G_m^{(\lambda)}(x)\,e^{-ip\cdot x}
    \end{equation}
and possesses the discrete symmetry properties
\begin{align}
    L^{(\lambda)}_{m,M}(-p,x) &= L^{(\lambda)}_{m,M}(p,-x),
\\
L^{(\lambda)}_{M,m}(p,x) &= e^{-ip\cdot x}L^{(\lambda)}_{m,M}(-p,x).
\end{align}

Starting from the momentum-space representation 
\be\la{eq:LdefMom}
L^{(\lambda)}_{m,M}(p,x) = \int \frac{d^dq}{(2\pi)^d} \frac{e^{-iqx}}{(q^2+M^2)[(q-p)^2+m^2]}
\ee
and using the Feynman-parameter representation
\ba
\frac{1}{(q^2+M^2)[(q-p)^2+m^2]} 
&=& \int_0^1 d\alpha \frac{1}{((1-\alpha) (q^2+M^2) + \alpha [(q-p)^2+m^2])^2}
\nonumber\\ &=& \int_0^1 d\alpha\; \frac{1}{((q- \alpha p)^2 +  \alpha(1-\alpha)p^2 + (1-\alpha)M^2 + \alpha m^2)^2},
\nonumber
\ea
one obtains, after shifting the integration variable $q\to q+\alpha p$, %  and swapping $\alpha \to 1-\alpha$,
\ba
L^{(\lambda)}_{m,M}(p,x) &=& \int_0^1 d\alpha \; e^{-i\alpha x\cdot p}\; \Big(\!-\frac{\partial}{\partial\mu^2}\Big) G_{\mu}^{(\lambda)}(x)\Big|_{\mu=\mu(\alpha)},
\\ \mu(\alpha) &=& \sqrt{\alpha(1-\alpha) p^2+(1-\alpha)M^2 +\alpha m^2}. \phantom{\bigg|}
\ea
Applying Eq.\ (\ref{eq:Gmderiv}), we arrive at the integral representation
\be\la{eq:Lfinal}
L^{(\lambda)}_{m,M}(p,x) = \frac{1}{4\pi}\int_0^1 d\alpha \; e^{-i\alpha x\cdot p}\; G_{\mu(\alpha)}^{(\lambda-1)}(x).
\ee
A simple but physically interesting case is $M=m$ and $p^2=0$, in which case $\mu(\alpha)=m$ is constant and we have the explicit result
\be\la{eq:Llightlike}
L^{(\lambda)}_{m,m}(p,x)  = \frac{i}{4\pi  p\cdot x} (e^{-i  p\cdot x} -1 ) \, G_{m}^{(\lambda-1)}(x), \qquad (p^2=0).
\ee
These kinematics correspond to the amplitude for a particle of mass $m$ to propagate from the origin to point $x$, emitting an on-shell massless particle on the way. It is proportional to the self-convolution of the propagator, see Eq.\ (\ref{eq:GmStarGm}), the proportionality constant being the dimensionless factor $(e^{-i  p\cdot x} -1 ) {i}/({p\cdot x}) $.

Another tractable case, considered below, is $M=0$ and $p^2=-m^2$, since $\mu(\alpha)=\alpha m$ then becomes linear in the Feynman parameter.

\subsection{Gegenbauer polynomial expansion}

From Eq.\ (\ref{eq:Lfinal}), using the expansion of the plane wave Eq.\ (\ref{eq:eipxGg}), the expansion in Gegenbauer polynomials yields
\ba
L^{(\lambda)}_{m,M}(p,x) &=& \sum_{n\geq 0 } l_n^{(\lambda);{m,M}}(p^2,x^2) \; C_n^{(\lambda)}(\hat p\cdot \hat x),
\\
l_n^{(\lambda);{m,M}}(p^2,x^2) &=&  \frac{2^\lambda \Gamma(\lambda)(\lambda+n) (-i)^n}{4\pi\,|x|^\lambda|p|^\lambda}\int_0^1  
\frac{d\alpha}{\alpha^\lambda}\, J_{\lambda+n}(\alpha|x||p|)  G_{\mu(\alpha)}^{(\lambda-1)}(x).
\la{eq:lncoeffFeyn}
\ea
The power series of $J_{\lambda+n}(\alpha |p||x|)/|p|^\lambda$ consists of $|p|^n$ times a series in $p^2$, and it is straightforward to analytically continue the function $l_n^{(\lambda);{m,M}}(p^2,x^2)$ to negative values of $p^2$, down to the on-shell value $p^2=-m^2$. Indeed
a tractable case emerges for\footnote{The restriction on $\lambda$ is necessary to avoid an infrared divergence.}
\begin{equation}
    M=0, \qquad p^2+m^2=0, \qquad \lambda>\frac{1}{2}\;.
\end{equation}
The function $L^{(\lambda)}_{m,0}(p,x)_{p^2+m^2=0}$ is the amplitude for a massive particle to propagate from the origin and emerge on-shell, having emitted a massless boson that reaches point $x$. This function appears in the coordinate-space approach to computing the hadronic light-by-light contribution to the muon $(g-2)$~\cite{Asmussen:2016lse,Blum:2017cer,Asmussen:2022oql}. It has also been used to rederive Schwinger's O($\alpha$) result for the $(g-2)$ of a lepton~\cite{BachelorThesisStuhlfauth}.
Starting from Eq.\ (\ref{eq:lncoeffFeyn}), we use Lommel~1 (see Eq.\ (\ref{eq:lom1KIb})) and obtain\footnote{A prime on a function denotes the derivative with respect to its argument.}
\ba\label{eq:ell_res1}
l_n^{(\lambda);{m,0}}(-m^2,x^2) &=& \frac{ \Gamma(\lambda)(\lambda+n)}{4 \pi^{\lambda+1} m \,|x|^{2\lambda-1}  } 
\int_0^1 \frac{d\alpha}{\alpha} \; I_{\lambda+n}(\alpha m|x|) \,   K_{\lambda-1}(\alpha m|x|) 
\\ 
\nonumber \\ 
&=& \frac{ \Gamma(\lambda)(\lambda+n)}{4 \pi^{\lambda+1}  \,|x|^{2(\lambda-1)}  }
\frac{K_{\lambda-1}(m|x|) I_{\lambda+n}'(m|x|)  - I_{\lambda+n}(m|x|) K_{\lambda-1}'(m|x|) }{(n+1)(2\lambda+n-1)}.
\nonumber
\ea

An alternative expression to Eq.\ (\ref{eq:lncoeffFeyn}) for the coefficients $l_n^{(\lambda);{m,M}}(p^2,x^2) $ 
can be obtained by
starting from the definition Eq.\ (\ref{eq:Ldef}) of $L^{(\lambda)}_{m,M}(p,x)$,
using the multipole expansions of the exponential function and of the propagator $G_M(x-z)$, and exploiting the orthogonality relation (\ref{eq:CnOrthog}) of the Gegenbauer polynomials. We arrive at
\ba\la{eq:Lexpan}
 l_n^{(\lambda);{m,M}}(p^2,x^2) &=&
 \frac{2^\lambda \Gamma(\lambda) (\lambda+n)(-i)^n }{|p|^\lambda|x|^\lambda}
\\ && \times \Big\{ K_{\lambda+n}(M|x|) \int_0^{|x|} dz\,z\,G_m^{(\lambda)}(z)\,I_{\lambda+n}(Mz)\,J_{\lambda+n}(|p|z)
\nonumber\\ && + I_{\lambda+n}(M|x|) \int_{|x|}^\infty dz \,z\, G_m^{(\lambda)}(z)\,K_{\lambda+n}(Mz)\,J_{\lambda+n}(|p|z) \Big\}.
\nonumber
\ea
It is interesting to compare this representation of the coefficients to that in Eq.\ (\ref{eq:lncoeffFeyn}), in which the integrand consists of two Bessel functions. In Eq.\ (\ref{eq:Lexpan}), three Bessel functions appear under the integral, albeit with an argument linear in the integration variable.

We now focus on the case $M=0$, which we obtain via a limit from Eq.\ (\ref{eq:Lexpan}). We assume $\lambda>1/2$ throughout this subsection.
Using $I_\lambda(az) K_\lambda(z)\, {\longrightarrow}\, a^\lambda/(2\lambda)$ in the limit $z\to0$, we arrive at
\ba 
l_n^{(\lambda);{m,0}}(p^2,x^2) 
  &=& \frac{\Gamma(\lambda)}{4\pi}
\Big(\frac{m}{\pi \,|p||x|}\Big)^\lambda 
(-i)^n 
\\ && \times\Big\{  \frac{1}{ |x|^{\lambda+n}} \int_0^{|x|} dz\,z^{n+1}\,  K_\lambda(mz)\,J_{\lambda+n}(|p|z)
\nonumber \\ && +  |x|^{\lambda+n}  \int_{|x|}^\infty dz \, \frac{1}{z^{2\lambda+n-1}}\,
K_\lambda(mz)  \,J_{\lambda+n}(|p|z) \Big\}.
\nonumber
\ea
If we take $p = im \hat p$, with $\hat p$ a Euclidean unit vector, and using $J_\lambda(iz) = i^\lambda\, I_\lambda(z)$, one finds 
\ba
 l^{(\lambda);{m,0}}_n(-m^2,x^2) &=&
\frac{\Gamma(\lambda)}{4\pi^{\lambda+1}\,|x|^\lambda} 
\Big\{  \frac{1}{ |x|^{\lambda+n}}\, {\cal I}_1(\lambda,n,m,|x|)
 +  |x|^{\lambda+n}\,  {\cal I}_2(\lambda,n,m,|x|,\infty) \Big\}, \qquad \quad 
\ea 
where the same integrals ${\cal I}_1$ and ${\cal I}_2$ appear as in the calculation of the coordinate-space three-point function $W(x,y)$; see Eqs.\ (\ref{eq:I1def}--\ref{eq:I2def}). Using the explicit expressions for ${\cal I}_1$ and ${\cal I}_2$ given in Eqs.\ (\ref{eq:I1res}--\ref{eq:I2res}), a result is obtained for the $l_n^{(\lambda);{m,0}}(-m^2,x^2)$ that is equivalent to that of Eq.\ (\ref{eq:ell_res1}).

\section{Extension to finite temperature\label{sec:finiteT}}

The scalar propagator at finite temperature can be written in several useful forms,
\begin{align}
    G_m^{(T,\lambda)}(x) 
    &\equiv  T\sum_{p_0} \int \frac{d^{d-1}p}{(2\pi)^{d-1}}\, \frac{e^{ip\cdot x}}{p^2+m^2}
 \\ \label{eq:GmT1}  &= T\sum_{p_0} e^{ip_0x_0}\,G_{\sqrt{p_0^2+m^2}}^{(\lambda-1/2)}(\vec x)
   \\ &= \sum_{n_0\in\mathbb{Z}} G_m^{(\lambda)}(x+\beta n_0\hat e_0),
   \label{eq:GmT2} 
\end{align}
where the sum over $p_0$ extends over the Matsubara modes, $p_0=2\pi Tn$, $n\in\mathbb{Z}$; $\beta=1/T$ is the inverse temperature and $\hat e_0$ is a unit vector pointing in the Euclidean time direction.

The relations (\ref{eq:GmStarGM})-(\ref{eq:GmStarGm}) carry over straightforwardly to finite temperature. Using the shorthand $\int_\beta d^dy\equiv\int_0^\beta dy_0\int d^{d-1}y$,
\begin{align}
 (G^{(T,\lambda)}_m *  G_M^{(T,\lambda)})(x) &\equiv  \int_\beta d^{d}y\; G^{(T,\lambda)}_m(x-y) G^{(T,\lambda)}_M(y) 
\nonumber \\ &= 
    \frac{1}{M^2-m^2} \left(G^{(T,\lambda)}_m(x) - G^{(T,\lambda)}_M(x) \right),
    \\
\label{eq:GmStarGmT}
  (G^{(T,\lambda)}_m *  G^{(T,\lambda)}_m)(x) 
 &\equiv \int_\beta d^{d}y\; G^{(T,\lambda)}_m(x-y) G^{(T,\lambda)}_m(y) 
\nonumber \\ &= -\frac{\partial}{\partial m^2}G^{(T,\lambda)}_m(x)
  = \frac{1}{4\pi} G^{(T,\lambda-1)}_m(x)\,.  
\end{align}

As an application, we consider the finite-temperature version of the one-loop correction to the scalar propagator (\ref{eq:2pt1loop_f1}).
Here we generalize the expression to a wider set of masses,
\begin{align}
    \label{eq:2pt1loop_f1T}
   & \delta_2 G_{m_1,m_2,m}^{(T,\lambda)}(x)
\\ \equiv& \;\int_\beta d^dy \int_\beta d^dz\; G_m^{(T,\lambda)}(x-y) G_{m_1}^{(T,\lambda)}(y-z)G_{m_2}^{(T,\lambda)}(y-z)G_m^{(T,\lambda)}(z)
\nonumber\\ =& \frac{1}{4\pi}\int_\beta d^dw\; G^{(T,\lambda)}_{m_1}(w)\, G^{(T,\lambda)}_{m_2}(w)\,  G^{(T,\lambda-1)}_m(x-w)
\label{eq:2pt1loop_f4T}
\\ =& \frac{1}{4\pi}\sum_{n_1,n_2,n\in\mathbb{Z}}\int_\beta d^dw\; G^{(\lambda)}_{m_1}(w+n_1\beta \hat e_0)\, G^{(\lambda)}_{m_2}(w+n_2\beta \hat e_0)\,  G^{(\lambda-1)}_m(x-w+n\beta \hat e_0)
\\ =& \frac{1}{4\pi}\sum_{\nu_1,\nu_2\in\mathbb{Z}}\int d^dw\; G^{(\lambda)}_{m_1}(w+\nu_1\beta \hat e_0)\, G^{(\lambda)}_{m_2}(w+\nu_2\beta \hat e_0)\,  G^{(\lambda-1)}_m(x-w)
\\ =& \frac{1}{4\pi}\sum_{\nu_1,\nu_2\in\mathbb{Z}}\int d^dv\; G^{(\lambda)}_{m_1}(x-v+\nu_1\beta \hat e_0)\, G^{(\lambda)}_{m_2}(x-v+\nu_2\beta \hat e_0)\,  G^{(\lambda-1)}_m(v).
\end{align}
In order to reach the form (\ref{eq:2pt1loop_f4T}), we have exploited translation invariance as for reaching Eq.\ (\ref{eq:2pt1loop_f2}), as well as the identity (\ref{eq:GmStarGmT}).
In the penultimate step, we have suppressed an overall shift of the arguments of the propagators by an integer multiple of $\beta$ in the time direction (setting $n_i=n+\nu_i$), and compensated by integrating $w_0$ over the entire real axis, rather than restricting it to the interval $0$ to $\beta$. In the last step we have made the change of variables $v=x-w$.

For QED-like kinematics, we arrive at the result
\begin{align}\label{eq:2pt1loop_f1Tresult}
    & \delta_2 G_{m,0,m}^{(T,\lambda)}(x)
 = \frac{1}{4\pi}\sum_{\nu_1,\nu_2\in\mathbb{Z}}
\widetilde W(x+\nu_1 \beta \hat e_0, x+\nu_2 \beta \hat e_0),
\end{align}
 where the expansion of the three-point function $\widetilde W(x,y)$ in Gegenbauer polynomials $C^{(\lambda)}_n(\hat x\cdot y)$ has been obtained in section \ref{sec:Wtilde}, see Eqs.\ (\ref{eq:Wtilde}) and (\ref{eq:wntilde_final}).
 We note that, in momentum space, the one-loop electron self-energy is well-known~\cite{Weldon:1982bn,Klimov:1982bv,Laine:2016hma}.

\section{Conclusion\label{sec:concl}}

We have shown that, in the Gegenbauer-polynomial  position-space approach to perturbative, dimensionally regularized quantum field theory, the radial integrals are still largely tractable in QED-like kinematics, i.e.\ with one mass scale present. To this end, we reviewed and extended classic results on modified Bessel functions, deriving conditions under which a certain class of antiderivatives takes the same form as the function under the integral.

As an illustration of calculations performed in position space, we computed the leading-order three-point function defined in Eq.\ (\ref{eq:Wwn}), with the result given by Eq.\ (\ref{eq:w_nFINAL}), as well as the one-loop correction to a two-point function, defined by Eq.\ (\ref{eq:d2GmDef}) and given in its final form in Eq.\ (\ref{eq:d2GmFINALform1}), keeping the number of spacetime dimensions $d=2\lambda+2$ general. Similarly, we have provided an expression for the position-space two-point function at non-zero temperature, to one-loop order (Eqs.\ (\ref{eq:2pt1loop_f1T}) and (\ref{eq:2pt1loop_f1Tresult})). In this context, we have illustrated that the underlying O($d$) spacetime symmetry can still be exploited, despite its breaking through the thermal boundary conditions in time, the price to pay being remaining sums over the winding numbers of propagators.

In position-space perturbation theory, one can define and compute re-usable `building blocks' such as the coordinate-space three-point function illustrated by Fig.\ \ref{fig:3ptFct} or the mixed coordinate/momentum space function defined by the diagram of Fig.\ \ref{fig:3ptmixed}. 
Once a suitable representation has been found for them, these building blocks can be used in a more extended diagram, reducing the actual number of integrals to be performed for such a diagram with $V$ vertices to a number smaller than $V-1$.
Obviously, as one moves to building blocks with additional external vertices, i.e.\ ones that are not integrated over, they become functions of more and more variables and thereby more complex. 

In this article, we have focused on the direct evaluation of the required angular and radial integrals. However, as stated in several instances throughout the manuscript, most of these integrals amount to solving an inhomogeneous second-order linear differential equation with the help of its Green's function. It is conceivable that other methods could be successfully applied to solving these differential equations. In particular, if the form of the answer is foreseeable on physics grounds, e.g.\ on the basis of a dispersive representation, making an ansatz with coefficients to be adjusted is a promising approach.

\section*{Acknowledgements}We thank Julian Parrino for sharing his experience from  his earlier work on coordinate-space perturbation theory in the context of the muon $(g-2)$~\cite{ParrinoMasterArbeit}. This work is supported by the Deutsche Forschungsgemeinschaft (German Research Foundation, DFG) through the Collaborative Research Center~1660 “Hadrons and Nuclei as Discovery Tools” and through the Cluster of Excellence “Precision Physics, Fundamental Interactions and Structure of Matter” (PRISMA+ EXC 2118/1), funded within the German Excellence strategy (Project No.\ 390831469). 

\appendix

\section{Two-point function at one loop order for $\lambda=1$\label{eq:lda1}}

Here we give the additional calculation for the quantity $\delta_2 G_m^{(\lambda)}(x)$ for $\lambda = 1$. Starting from Eq.\ (\ref{eq:2pt1loop_f4}), 
one gets instead
\begin{equation}
	\Bigl(\delta_2 G_m^{(1)}(x)\Bigr)_1 = \frac{m}{2^7 \pi^6} \int \mathrm{d}^4w\, \frac{K_1(m|w|)}{|w|^3} K_0(m|x-w|)
\end{equation}
and can apply the Gegenbauer expansion in the form (cf. appendix \ref{appdx_k0})
\begin{equation}
	K_0(m|x-w|) = \sum_{n=0}^\infty U_n(\hat{x} \cdot \hat{w}) \begin{cases}K_n(m|x|) I_n(m|w|) - K_{n+2}(m|x|) I_{n+2}(m|w|), &|x| \ge |w|,\\K_n(m|w|) I_n(m|x|) - K_{n+2}(m|w|) I_{n+2}(m|x|), &|x| < |w|.\end{cases}
\end{equation}
This gives, on exploitation of the orthogonality of the Chebyshev polynomials,
\begin{align}
	\Bigl(\delta_2 G_m^{(1)}(x)\Bigr)_1 = \frac{m}{2^6 \pi^4} \Biggl\{&K_0(m|x|) \int_0^{|x|} \mathrm{d}|w|\, K_1(m|w|) I_0(m|w|) + \nonumber\\
	&I_0(m|x|) \int_{|x|}^\infty	 \mathrm{d}|w|\, K_1(m|w|) K_0(m|w|) - \nonumber\\
	&K_2(m|x|) \int_0^{|x|} \mathrm{d}|w|\, K_1(m|w|) I_2(m|w|) - \nonumber\\
	&I_2(m|x|) \int_{|x|}^\infty \mathrm{d}|w|\, K_1(m|w|) K_2(m|w|)\Biggr\}.
\end{align}
So the general case reduces to this when one puts $\lambda = 1$ in eq. \eqref{eq:se_conv_general_ints}.

\subsection{Solving the integrals for $\lambda=1$}
We will now derive antiderivatives for the integrals in four dimensions. First, we use the substitution $u = m|w|$ in each, we thus arrive at
\begin{align}
	\Bigl(\delta_2 G_m^{(1)}(x)\Bigr)_1 = \frac{1}{2^6 \pi^4} \Biggl\{&K_0(m|x|) \int_0^{m|x|} \mathrm{d}u\, K_1(u) I_0(u) + \nonumber\\
	&I_0(m|x|) \int_{m|x|}^\infty	 \mathrm{d}u\, K_1(u) K_0(u) - \nonumber\\
	&K_2(m|x|) \int_0^{m|x|} \mathrm{d}u\, K_1(u) I_2(u) - \nonumber\\
	&I_2(m|x|) \int_{m|x|}^\infty \mathrm{d}u\, K_1(u) K_2(u)\Biggr\}.
\end{align}

For the first of the four integrals use $K_1(z) = - K_0'(z)$ and integration by parts,
\begin{align}
	\int \mathrm{d}u\, K_1(u) I_0(u) &= -\int \mathrm{d}u\, K_0'(u) I_0(u) \nonumber\\
	&= -K_0(u) I_0(u) + \int \mathrm{d}u\, K_0(u) I_1(u).
\end{align}
Now on using the identity $K_0(z) I_1(z) + K_1(z) I_0(z) = 1/z$ we arrive at
\begin{equation}
	\int \mathrm{d}u\, K_1(u) I_0(u) = -K_0(u) I_0(u) + \int \mathrm{d}u\, \frac{1}{u} - \int \mathrm{d}u\, K_1(u) I_0(u),
\end{equation}
which can be rearranged to
\begin{equation}
	2 \int \mathrm{d}u\, K_1(u) I_0(u) = -K_0(u) I_0(u) + \log |u|.	
\end{equation}
This gives the first antiderivative.

For the third we first use the integral recursion relation in Lommel 2 to rewrite the integral as
\begin{equation}
	\int \mathrm{d}u\, K_1(u) I_2(u) = -\frac{u}{2} \Bigl(K_0(u) I_1(u) + K_1(u) I_2(u)\Bigr) + \int \mathrm{d}u\, K_0(u) I_1(u).
\end{equation}
Next, we again make use of the identity $K_0(z) I_1(z) + K_1(z) I_0(z) = 1/z$,
\begin{equation}
	\int \mathrm{d}u\, K_1(u) I_2(u) = -\frac{u}{2} \Bigl(K_0(u) I_1(u) + K_1(u) I_2(u)\Bigr) + \log |u| - \int \mathrm{d}u\, K_1(u) I_0(u),
\end{equation}
and then can insert the result from above for the first antiderivative,
\begin{align}
	\int \mathrm{d}u\, K_1(u) I_2(u) = -\frac{u}{2} \Bigl(K_0(u) I_1(u) + K_1(u) I_2(u)\Bigr) + \frac{1}{2}\Bigl(K_0(u) I_0(u) + \log |u|\Bigr).
\end{align}
On may now still use the identity $K_0(u) I_1(u) + K_1(u) I_0(u) = 1/z$ and the recursion relation for the Bessel function to simplify this, which yields
\begin{equation}
	\int \mathrm{d}u\, K_1(u) I_2(u) = \frac{1}{2} \Bigl\{\log |u| - 1 + 2 I_1(u) K_1(u) + K_0(u) I_0(u)\Bigr\}.
\end{equation}

The second antiderivative can again be computed by the same procedure used for the first: it follows that
\begin{equation}
	\int \mathrm{d}u\, K_1(u) K_0(u) = -\frac{1}{2} K_0^2(u).
\end{equation}

Finally, the fourth antiderivative can be obtained again by an application of Lommel 2,
\begin{equation}
	\int \mathrm{d}u\, K_1(u) K_2(u) = -\int \mathrm{d}u\, K_0(u) K_1(u) + \frac{u}{2} \Bigl(K_0(u) K_1(u) - K_1(u) K_2(u)\Bigr).
\end{equation}
Using the above result for the third antiderivative finally yields
\begin{equation}
	\int \mathrm{d}u\, K_1(u) K_2(u) = \frac{1}{2} K_0^2(u) - K_1^2(u).
\end{equation}

\subsubsection{Evaluation of the definite integrals}
Now one may plug in the integral boundaries into the antiderivatives. For the last three integrals this is straight-forward, since they all converge, as we will now see. For the second integral we have a contribution of
\begin{equation}
	\Bigl(\delta_2 G_m^{(1)}(x)\Bigr)_{1,2} = \frac{1}{2^6 \pi^4} I_0(m|x|) \Bigl[-\frac{1}{2} K_0^2(u)\Bigr]_{u=m|x|}^\infty.
\end{equation}
Since $K_\nu(z) \sim \sqrt{\frac{\pi}{2z}} e^{-z}$ for large $z$ the upper boundary does not contribute and we get
\begin{equation}
	\Bigl(\delta_2 G_m^{(1)}(x)\Bigr)_{1,2} = \frac{1}{2^7 \pi^4} I_0(m|x|) K_0^2(m|x|).
\end{equation}

For the contribution from the fourth integral we have
\begin{equation}
	\Bigl(\delta_2 G_m^{(1)}(x)\Bigr)_{1,4} = -\frac{1}{2^6 \pi^4} I_2(m|x|) \Bigl[\frac{1}{2} K_0^2(u) - K_1^2(u)\Bigr]_{u=m|x|}^\infty.
\end{equation}
Again, the upper boundary does not contribute because of the asymptotic behaviour of the modified Bessel function of the second kind. Hence,
\begin{equation}
	\Bigl(\delta_2 G_m^{(1)}(x)\Bigr)_{1,4} = \frac{1}{2^6 \pi^4} I_2(m|x|) \Bigl[\frac{1}{2} K_0^2(m|x|) - K_1^2(m|x|)\Bigr].
\end{equation}

For the contribution of the third integral it is not as clear that it converges. But observe that from the limiting form of the modified Bessel functions for small arguments, one can deduce that for $\nu > 0$ and $\mu > 0$ \cite{nist}
\begin{equation}
	K_\mu(z) I_\nu(z) \sim \frac{1}{2} \Gamma(\mu) \left(\frac{2}{z}\right)^\mu \left(\frac{z}{2}\right)^\nu \frac{1}{\Gamma(\nu+1)} = \frac{1}{2} \frac{\Gamma(\mu)}{\Gamma(\nu+1)} \left(\frac{2}{z}\right)^{\mu-\nu},
\end{equation}
whereas for $\mu=0$ we have \cite{nist}
\begin{equation}
	K_0(z) I_\nu(z) \sim -\log z \left(\frac{z}{2}\right)^\nu \frac{1}{\Gamma(\nu+1)}.
\end{equation}
and more precisely, by using the series expansions, \cite{Watson}
\begin{equation}
	K_0(z) I_0(z) = (-\gamma_E + \log 2 - \log z) + \mathcal{O}(z).
\end{equation}
Then, we find that
\begin{align}
	\lim_{u \downarrow 0} \frac{1}{2} &\Bigl(\log u - 1 + 2 I_1(u) K_1(u) + K_0(u) I_0(u)\Bigr) \nonumber\\
	&= \lim_{u \downarrow 0} \frac{1}{2} \Bigl(\log u - 1 + 1 - \gamma_E + \log 2 - \log u\Bigr) \nonumber\\
	&= \frac{\log 2 - \gamma_E}{2}.
\end{align}
Now one may compute the contribution of the third integral,
\begin{align}
	\Bigl(\delta_2 G_m^{(1)}(x)\Bigr)_{1,3} &= -\frac{1}{2^6 \pi^4} K_2(m|x|) \frac{1}{2} \Bigl[\log |u| - 1 + 2 I_1(u) K_1(u) + K_0(u) I_0(u)\Bigr]_{u=0}^{m|x|} \nonumber\\
	&= \frac{1}{2^7 \pi^4} K_2(m|x|) \Bigl\{1 + \log 2 - \gamma_E - \log(m|x|) \nonumber\\
	&\qquad - 2 I_1(m|x|) K_1(m|x|) - K_0(m|x|) I_0(m|x|)\Bigr\}.
\end{align}

Finally, the first integral does not converge. We may still plug in the boundaries and treat the infinite limit as an (improper) constant. This gives
\begin{align}
	\Bigl(\delta_2 G_m^{(1)}(x)\Bigr)_{1,1} &= \frac{1}{2^7 \pi^4} K_0(m|x|) \Biggl\{\log(m|x|) - K_0(m|x|) I_0(m|x|) \nonumber\\
	&\qquad - \gamma_E + \log 2 - \lim_{u \downarrow 0} \Bigl(2\log u \Bigr)\Biggr\}.
\end{align}
The full result is then given by
\begin{equation}
	\Bigl(\delta_2 G_m^{(1)}(x)\Bigr)_{1} = \Bigl(\delta_2 G_m^{(1)}(x)\Bigr)_{1,1} + \Bigl(\delta_2 G_m^{(1)}(x)\Bigr)_{1,2} + \Bigl(\delta_2 G_m^{(1)}(x)\Bigr)_{1,3} + \Bigl(\delta_2 G_m^{(1)}(x)\Bigr)_{1,4}.
\end{equation}

Now, one may analyse the infrared behaviour of the result. To do so, one uses the limiting forms for large arguments of the modified Bessel functions, which are given by \cite{nist}
\[
	K_\nu(z) \sim \sqrt{\frac{\pi}{2z}} e^{-z}, \qquad I_\nu(z) \sim \frac{e^z}{\sqrt{2\pi z}}, \qquad (z \to \infty).
\]
This gives
\begin{equation}
	\Bigl(\delta_2 G_m^{(1)}(x)\Bigr)_{1} \sim \frac{1}{2^6 \pi^4} \sqrt{\frac{\pi}{2}} \Biggl\{\Bigl(\frac{1}{2} + \log 2 - \gamma_E - \lim_{u \downarrow 0} \log u\Bigr) \frac{e^{-m|x|}}{\sqrt{m|x|}} - \frac{e^{-m|x|}}{(m|x|)^{3/2}}\Biggr\}.
\end{equation}

Note that from a mass insertion $\mathcal{L}_E = \delta m^2 \frac{\phi^2}{2}$ into the Lagrangian density of our toy model we obtain a counterterm contribution
\begin{equation}
	-\delta m^2 \int_y G_m^{(\lambda)}(x-y) G_m^{(\lambda)}(y) = -\delta m^2 \frac{G_m^{(\lambda-1)}(x)}{4\pi},	
\end{equation}
where the convolution identity (eq. \eqref{eq:GmStarGm}) was used. It is therefore possible to cancel this divergence and obtain the correct infrared behaviour by setting $\delta m^2$ appropriately.

\section{The two-point function at one loop order for $\lambda>1$\label{sec:ldageq1}}

In the following, we present the calculation of the definite integrals from the two-loop function at one loop order in the case $\lambda > 1$. We refer to the integrals derived for general $\lambda$ in section \ref{dfnt_integrals}. We especially present a discussion of the infinities arising here.

The first integral was already ultraviolet-divergent in the four-dimensional case, therefore we expect it to remain so at $\lambda > 1$ \emph{a fortiori}. Using the antiderivative, we obtain
\begin{align}
	\int_0^{m|x|} \mathrm{d}u\, u^{2-2\lambda} K_\lambda(u) I_{\lambda-1}(u) &= -\frac{(m|x|)^{2-2\lambda}}{2} \Bigl(\frac{1}{2(\lambda-1)} + K_{\lambda-1}(m|x|) I_{\lambda-1}(m|x|)\Bigr) \nonumber\\
	&\qquad+ \lim_{u \downarrow 0} \frac{u^{2-2\lambda}}{2} \Bigl(\frac{1}{2(\lambda-1)} + K_{\lambda-1}(u) I_{\lambda-1}(u)\Bigr).
\end{align}
If we investigate the limit for $\lambda \ne 1$ we find that by inserting the series expansions of the Bessel functions for small arguments,
\begin{equation}
	\lim_{u \downarrow 0} \Bigl(\frac{1}{2(\lambda-1)} + K_{\lambda-1}(u) I_{\lambda-1}(u)\Bigr) = \lim_{u \downarrow 0} \Bigl(\frac{1}{2(\lambda-1)} + \frac{1}{2(\lambda-1)}\Bigr) = \frac{1}{\lambda-1}
\end{equation}
is a constant. Hence,
\begin{equation}
	\lim_{u \downarrow 0} \frac{u^{2-2\lambda}}{2} \Bigl(\frac{1}{2(\lambda-1)} + K_{\lambda-1}(u) I_{\lambda-1}(u)\Bigr) = \infty,
\end{equation}
for $\lambda > 1$ and this integral also diverges in this case.

The most interesting integral is the third, since we will see new behaviour for $\lambda \ne 1$ compared to the four-dimensional case. Let us first insert the boundaries. One obtains
\begin{align}
	\int_0^{m|x|} &\mathrm{d}u\, u^{2-2\lambda} K_\lambda(u) I_{\lambda+1}(u) = \nonumber\\
	& \frac{(m|x|)^{2-2\lambda}}{2(1-2\lambda)} \Biggl(\frac{2\lambda-1}{2\lambda-2} - 2\lambda K_\lambda(m|x|) I_\lambda(m|x|) - K_{\lambda-1}(m|x|) I_{\lambda-1}(m|x|)\Biggr) \nonumber\\
	&\qquad - \lim_{u \downarrow 0} \frac{u^{2-2\lambda}}{2(1-2\lambda)} \Biggl(\frac{2\lambda-1}{2\lambda-2} - 2\lambda K_\lambda(u) I_\lambda(u) - K_{\lambda-1}(u) I_{\lambda-1}(u)\Biggr).
\end{align}

Looking at the limit for $\lambda > 1$, one may first compute
\begin{align}
	\lim_{u \downarrow 0} \Biggl(\frac{2\lambda-1}{2\lambda-2} &- 2\lambda K_\lambda(u) I_\lambda(u) - K_{\lambda-1}(u) I_{\lambda-1}(u)\Biggr) = \nonumber\\
	&\lim_{u \downarrow 0} \Biggl(\frac{2\lambda-1}{2\lambda-2} - 2\lambda \frac{1}{2\lambda} - \frac{1}{2(\lambda-1)}\Biggr) = 0.
\end{align}
This is again done by inserting the series expansion for the modified Bessel functions for small argument. An application of L'H\^{o}pital's rule now shows the following,
\begin{equation}\label{eq:se:lhospital}
	\lim_{u \downarrow 0} u^{2-2\lambda} \Bigl(\dots\Bigr) = \lim_{u \downarrow 0} \frac{\Bigl(\dots\Bigr)}{u^{2\lambda-2}} = \lim_{u \downarrow 0} \frac{\Bigl(\dots\Bigr)'}{2(\lambda-1)u^{2\lambda-3}}.
\end{equation}
Since by explicit calculation we find that 
\begin{equation}
	\Bigl(\dots\Bigr)' = (-2\lambda + 1) K_\lambda(u) I_{\lambda-1}(u) - K_{\lambda-1}(u) I_{\lambda-2}(u) + 2\lambda K_{\lambda+1}(u) I_\lambda(u),
\end{equation}
where for $1 < \lambda < 3/2$ we get by looking at the series expansion of the Bessel functions for small argument
\begin{equation}
	\lim_{u \downarrow 0} \Bigl(\dots\Bigr)' = 0,	
\end{equation}
and $\lim_{u \downarrow 0} u^{2\lambda-3} > 0$ for $\lambda \le 3/2$, we conclude that the third integral is finite for $1 < \lambda < 3/2$. For the special case of $\lambda = 3/2$ we instead have to take into account the series expansion up to zeroth order, namely
\begin{equation}
	K_{1/2}(z) I_{-1/2}(z) = \frac{1}{z} - 1 + \mathcal{O}(z),
\end{equation}
where the zeroth order does not vanish for this particular case. Hence,
\begin{equation}
	\lim_{u \downarrow 0} \Bigl(\dots\Bigr)' = 1
\end{equation}
for $\lambda = 3/2$. Going back to eq. \eqref{eq:se:lhospital} this implies
\begin{equation}
	\lim_{u \downarrow 0} u^{2-2\lambda} \Bigl(\dots\Bigr) = 1. 
\end{equation}
Therefore, one can say that the integral remains finite	for $\lambda \le 3/2$. Let us now investigate the case $\lambda > 3/2$. Starting from eq. \eqref{eq:se:lhospital} and applying L'H\^{o}pital's rule again, one obtains
\begin{equation}\label{eq:se:lhospital2}
	\lim_{u \downarrow 0} u^{2-2\lambda} \Bigl(\dots\Bigr) = \lim_{u \downarrow 0} \frac{\Bigl(\dots\Bigr)''}{2(\lambda-1)(2\lambda-3) u^{2\lambda-4}}.
\end{equation}
Again by explicit calculation and usage of the series expansions for small argument, one gets that
\begin{equation}
	\lim_{u \downarrow 0} \Bigl(\dots\Bigr)'' = \text{const.}	
\end{equation}
for $\lambda > 2$, whereas
\begin{equation}
	\lim_{u \downarrow 0} \Bigl(\dots\Bigr)'' = \infty	
\end{equation}
for $3/2 < \lambda \le 2$. In the case of $\lambda = 2$ the divergence is of logarithmic nature. Thus, we can conclude that in all these cases (where $\lambda > 3/2$), by virtue of eq. \eqref{eq:se:lhospital2}, the third integral is infinite.

Since the form of the divergence is not the same for different values of $\lambda$, we do not state a final result for the function $\delta_2 G_m^{(\lambda)}$ here.

\section{Derivation of the Gegenbauer expansion for $K_0(|x-u|)$}\label{appdx_k0}

We start by taking Neumann's addition theorem (resp. a special case of Graf's addition theorem) in the form \cite{Watson}
\begin{equation}\label{eq:appdx_k0:grafKcos}
	K_0(\varpi) = \sum_{k=-\infty}^\infty K_{k}(Z) I_k(z) \cos(k\varphi),	
\end{equation}
where $\varpi = \sqrt{Z^2 + z^2 - 2Zz\cos\varphi}$ for complex values $Z, z, \varphi$.

Let us now adopt this to our situation and set $Z = |u|$ and $z = |x|$ to be positive values. Set $\varphi$ to be the angle between the vectors $u$ and $x$, so that $\cos\varphi = \hat{u} \cdot \hat{x}$. Then, by definition, $\varpi = |x-u|$. 

Now we can start to prove the actual formula in question. First, transform the summation, 
\begin{equation}
	K_0(|x-u|) = K_0(|u|) I_0(|x|) + 2\sum_{k=1}^\infty I_k(|x|) K_k(|u|) \cos(k\varphi).
\end{equation}
Then, introduce the Chebyshev polynomial of the first kind $T_n$ by the identity
\begin{equation}\label{eq:appdx_k0:Tndef}
	T_n(\cos\theta) = \cos(n\theta),	
\end{equation}
and combine this with the well-known interrelation between the Chebyshev polynomials,
\begin{equation}\label{eq:appdx_k0:TnUnrel}
	T_n(x) = \frac{1}{2} \Bigl(U_n(x) (1 + \delta_{n0}) - U_{n-2}(x)\Bigr).
\end{equation}
Here, the convention is used that the Chebyshev polynomial vanishes for negative orders. Upon insertion we get
\begin{align}
	K_0(|x-u|) &= \sum_{k=0}^\infty I_k(|x|) K_k(|u|) U_k(\cos\varphi) \nonumber\\
	&\qquad- \sum_{k=0}^\infty I_k(|x|) K_k(|u|) U_{k-2}(\cos\varphi).
\end{align}
The second summation actually starts at $k = 2$ and on an index shift this becomes
\begin{equation}
	K_0(|x-u|) = \sum_{k=0}^\infty U_k(\cos\varphi) \Bigl\{I_k(|x|) K_k(|u|) - I_{k+2}(|x|) K_{k+2}(|u|)\Bigr\}	.
\end{equation}

\bibliographystyle{JHEP}
\bibliography{references}
\end{document}